\documentclass[twocolumn,amsmath,amssymb,groupedaddress,superscriptaddress]{revtex4-1}
\usepackage[bookmarks=true,colorlinks,linkcolor=blue,urlcolor=blue,citecolor=blue]{hyperref}
\usepackage{graphics,color}
\usepackage{amsmath}
\usepackage[dvips]{graphicx}
\usepackage{float,graphicx}
\usepackage{subfigure,hyperref}
\usepackage{epstopdf}
\usepackage{epstopdf}
\usepackage{soul}
\usepackage{epstopdf}
\newcommand{\beq}{\begin{equation}}
\newcommand{\eeq}{\end{equation}}
\newcommand{\beqa}{\begin{eqnarray}}

\def\ME{\textmd{ME}}
\def\NE{\textmd{NE}}
\newcommand{\eeqa}{\end{eqnarray}}

\newcommand{\ud}{\,\mathrm{d}} 

\newcommand{\ket}[1]{\vert #1 \rangle} 

\def\be{\begin{equation}}

\def\ee{\end{equation}}

\def\eeff{\textmd{eff}}

\def\eeff{\textmd{eff}}

\def\ttot{\textmd{tot}}

\def\eeff{\textmd{eff}}
\def\Rre{\textmd{Re}}
\def\Iim{\textmd{Im}}

\def\Ttr{\textmd{Tr}}

\def\be{\begin{equation}}
\def\ee{\end{equation}}
\def\bea{\begin{eqnarray}}
\def\eea{\end{eqnarray}}

\usepackage{bm}
\usepackage{graphicx}
\begin{document}
\title[]{Dynamics of multiple atoms in one-dimensional fields}
\author{Carlo Cascio}
\affiliation{Department of Physics and Arnold Sommerfeld Center for Theoretical Physics, Ludwig-Maximilians-Universit{\"a}t M{\"u}nchen, Theresienstr. 37, 80333 Munich, Germany}
\author{Jad C. Halimeh}
\affiliation{Max Planck Institute for the Physics of Complex Systems, 01187 Dresden, Germany} \affiliation{Physik Department, Technische Universit\"at M\"unchen, 85747 Garching, Germany}
\author{Ian P. McCulloch}
\affiliation{ARC Centre of Excellence for Engineered Quantum Systems, School of Mathematics and Physics, The University
of Queensland, St. Lucia, QLD 4072, Australia}
\author{Alessio Recati}
\affiliation{INO-CNR BEC Center and Dipartimento di Fisica, Universit\`a di Trento, I-38123 Povo, Italy}
\affiliation{Trento Institute for Fundamental Physics and Applications, INFN, 38123, Trento, Italy}
\author{In\'es de Vega}
\affiliation{Department of Physics and Arnold Sommerfeld Center for Theoretical Physics, Ludwig-Maximilians-Universit{\"a}t M{\"u}nchen, Theresienstr. 37, 80333 Munich, Germany}
\date{\today}
\begin{abstract}
We analyze the dynamics of a set of two-level atoms coupled to the electromagnetic environment within a waveguide. This problem is often tackled by assuming a weak coupling between the atoms and the environment as well as the associated Markov approximation. We show that the accuracy of such an approximation may be more limited than in the single-atom case and also be strongly determined by the presence of collective effects produced by atom-atom interactions. To this aim, we solve the full problem with exact diagonalization and also the time-dependent density matrix renormalization group method, and compare the result to that obtained within a weak-coupling master equation and with the Dicke approximation. 
Finally, we study the dynamics of the entanglement within the system when considering several inter-atomic distances and atomic frequencies. 
\end{abstract}
\maketitle
\section{Introduction}\label{sec:Intro}
Artificial materials can be engineered to control the propagation of light as well as its absorption and scattering properties not only in the domain of classical optics (such as in photonic crystals \cite{john1987,yablonovitch1987} and metamaterials \cite{metabook2010}) but also at a quantum-mechanical level, as is the case in electromagnetically induced transparency (EIT) media \cite{fleischhauer2000,fleischhauer2005}. Besides its fundamental interest, understanding the interaction of light with a quantum-mechanical medium is primary to the development of nanoscale optoelectronic devices that may replace electronic devices in the future. One of the most interesting examples of this type of systems is a single or several emitters embedded in an optical waveguide  \cite{shen2009,zheng2013,dai2015,mascarenhas2015,fratini2015}, trapped near a photonic crystal waveguide \cite{douglas2013,tudela2014,caneva2015}, or coupled to propagating surface plasmons confined to a conducting nanowire \cite{chang2006,chang2007}. An advantage of this setup is that the three-dimensional electromagnetic field is reduced to a single dimension, which considerably simplifies the problem. This reduction can also be considered when an atomic ensemble is excited with a one-dimensional (1D) light input \cite{Hammerer2010a,zeuthen2011,manzoni2017}. Besides the quantum optics case, an atomic system interacting with a scalar bosonic field can be implemented with atoms having one internal level trapped by an optical lattice and coupled through Raman transitions to an untrapped level \cite{devega2008,navarrete2010}, or by considering quantum dots coupled to the excitations of a Bose-Einstein condensate \cite{recati2005,marino2017}. 

A single atom coupled to a 1D light field has been extensively analyzed  
\cite{chang2006,shen2007,zhou2008,shen2009,longo2010,roy2011,sanchez2017}, with the analysis having been more recently 
extended to two \cite{zheng2013,dai2015,mascarenhas2015,fratini2015} and more 
\cite{douglas2013,devega2014a,tudela2014,caneva2015} atoms. The case of 2D structured environments has been recently explored in \cite{tudela2017b,tudela2017}. Moreover, the dynamics of multiple atoms brings some of the most interesting applications and phenomena \cite{calajo2016}. For instance, two atoms embedded in a waveguide and 
having different resonant frequencies have been recently proposed as means to achieve 
unidirectional quantum transport of light \cite{dai2015,mascarenhas2015,fratini2015}, a property 
also known as rectification and that is relevant for achieving optical isolation of a circuit 
\cite{jalas2013}. In addition, a careful choice of the atomic frequencies with respect to the 
spectral density -- which is a function that characterizes their interaction with the environment -- allows one to produce long-range atom-atom interactions at low dissipation, 
which can be described with effective 
Hamiltonians \cite{devega2008,douglas2013,tudela2014,caneva2015,krinner2017}. 

Even if the electromagnetic field is reduced to one dimension, an accurate description of the atomic dynamics beyond the Born-Markov and the weak-coupling approximations is still challenging \cite{devega2015c}. Indeed, many previous studies rely on using a perturbative expansion with respect to the coupling between the system and its environment. For a single open system, such a weak-coupling approximation is inaccurate not only at strong couplings, but also when the system resonant frequencies are within a rapidly varying region of the spectral density. This occurs for instance in the vicinity of a band-gap edge. As a consequence, the initial environment state $\rho_\text{B}(0)$ is substantially perturbed and the environment may become significantly correlated with the open system, thereby hindering the validity of the weak-coupling approximation. 
Since many-body open systems can potentially produce a stronger perturbation of the environment state, a relevant question is whether in those cases the weak-coupling approximation is more inexact.

 \begin{figure}[!ht]
		\includegraphics[width=1\linewidth]{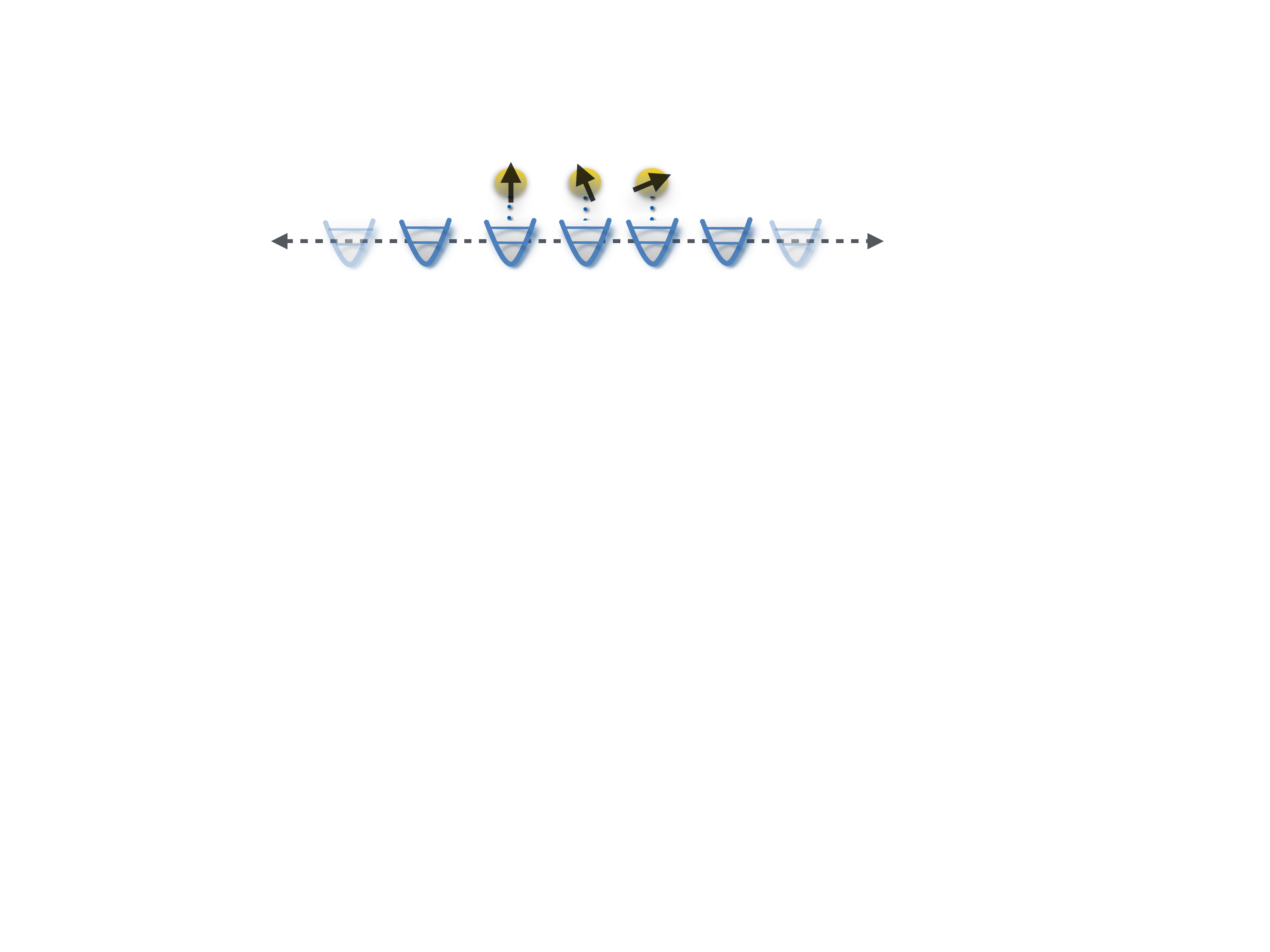}	
\caption{Schematic representation of the ladder-like structure of our system in~\eqref{DMRG_H}. The atoms, represented by the arrows above, are locally coupled to the 
transformed oscillators of the environment, conforming the infinite chain below.}
\label{thesystem}
\end{figure}
Here we analyze these issues by exploring the dynamics of an array of up to three atoms coupled to a 1D electromagnetic (e.m.) field. 
To this aim, we compare the weak coupling master equation with two different numerical methods that solve the evolution of the full system: exact diagonalization (ED) and time-dependent density matrix renormalization group (t-DMRG) \cite{white1992,white2004,scholl2005,scholl2005b}. 
Both methods offer a very accurate solution of the system dynamics: while ED computes the exact dynamics without truncation of the size of the Hilbert space, t-DMRG is based on truncating the Hilbert space, but it does so in a controlled way that allows to progressively reduce the truncation error until convergence to the exact result occurs.
We use ED when a single excitation is present in the problem since it is much faster than t-DMRG in this case, and we also consider it as an independent reference to check the t-DMRG implementation. 

However, we consider $t$-DMRG when there is more than one excitation in the problem since in this case it is much more efficient than ED. The $t$-DMRG approach we adopt here is based on matrix product states \cite{scholl2011,ian2007}, and the time-evolution routine is carried out using Krylov method \cite{LanczosTime} combined with time-evolving block decimation \cite{Vidal2003,Verstraete2004,Zwolak2004}. Furthermore, its use is facilitated by the fact that under certain additional conditions the system can be mapped into a ladder-like structure with local couplings \cite{devega2014a} (see Fig.~\ref{thesystem}), which in principle reduces the growth of entanglement and therefore the dimension of the Hilbert space explored during the evolution. More details on these methods are discussed in Appendix \ref{ed_dmrg_sec}.


The aim of the present work is two-fold: first of all, we characterize the performance of the 
second-order weak-coupling approximation in describing the evolution of an extended excited many-body open system.
Secondly, by using exact methods we study in detail the atomic dynamics depending on their relative distances and on whether their initial internal states are separable or entangled.

The paper is organized as follows: Sec.~\ref{periodic} introduces the model and the associated Hamiltonian. Following this, Sec.~\ref{WCA} analyzes the performance of the second-order weak-coupling approximation. Furthermore, in Sec.~\ref{variation} we use $t$-DMRG to study the system evolution for different initial conditions. By considering an initially entangled state for the atoms we study entanglement degradation, and, therefore, the suitability of those systems to serve as a quantum memory. We also analyze the effects of having different inter-atomic distances $L$ in the presence of cooperative phenomena in the emission. Finally, in Sec.~\ref{conclusions} we summarize the main conclusions of the paper.

\section{The model}
\label{periodic}

We consider $N$ identical two-level atoms located at fixed positions and coupled to a 1D 
coupled optical resonator waveguide (CROW), where the e.m field has a band-gap dispersion of the form \cite{yariv1999} 
\bea
\omega(k)=A+B\cos(kh_0).
\label{dispersion}
\eea
Here we are considering $M$ optical resonators separated by a distance $h_0$, and therefore the quasi-momenta are $k= \frac{2\pi q}{h_0M}$, 
with $q= -M/2 +1,...,M/2$. We also identify $\omega_\text{c}=A-B$ and
$\tilde{\omega}_\text{c}=A+B$ as the lower and upper bound of the propagating band, respectively. 
Considering the environment degrees of freedom in the position space, the full system Hamiltonian can be written as (see Appendix \ref{AppA} for details)
\bea
	H &=& A \sum_{j=-M/2+1}^{M/2} a_{j}^{\dagger} a_{j} + \frac{B}{2} \sum_{j=-M/2+1}^{M/2} \left( a^{\dagger}_{j+1}a_{j} + a^{\dagger}_{j}a_{j+1}\right) \cr
	&+&\sqrt{M}g \sum_{j=1}^{N} \left( a_{m(j)}\sigma_{j}^{+} + a_{m(j)}^{\dagger} \sigma_{j}\right)+H_\text{S},
	\label{DMRG_H}
\eea
where $a_{j}$ are the annihilation operators of the field at site $j$, and the label $m(j)$ refers to the actual atomic position of the $j$-th atom, which is a multiple 
of $h_0$. Moreover, introducing the atomic ground and excited states $|0\rangle_j$ and $|1\rangle_j$, respectively, for an atom at the position $r_j$  with an energy difference $\omega_\text{S}$, the free atomic Hamiltonian reads ($\hbar=1$) $H_\text{S}=\omega_\text{S}\sum_{j=1}^{N}\sigma_j^+\sigma_j$, where we have defined the spin ladder operators $\sigma_j^+=|1\rangle_j \langle 0|_j$. Moreover in the following we define $\Delta=\omega_\text{S} - \omega_\text{c}$,
the detuning between the atomic resonance frequency and the lower band-gap edge $\omega_\text{c}$. 

Thus, our system of atoms interacting with a 1D electromagnetic field is described by a Hamiltonian that can be schematically represented by the ladder-like structure displayed in Fig.~\ref{thesystem}. 
Aside from its physical relevance, this model is particularly convenient since standard numerical methods for studying strongly correlated 1D systems, such as $t$-DMRG, can be applied efficiently.
The present model can also describe the dynamics of impurities in 1D photonic crystals, provided that the detuning to any other band-gap edge is much larger than $\Delta$, so that the influence of other bands is 
negligible \cite{florescu2001,devega2005,douglas2013}.
Moreover, as also sketched in Fig.~\ref{thesystem}, we consider that the string of atoms is located well within the waveguide, such that boundary effects can be ignored.


To carry out the comparison between an exact treatment and a master equation we follow the time evolution of a few quantities related to the system population and consider an initial separable state. Within the exact treatment, we will further analyze the dynamics of the atomic entanglement as described by the concurrence as well as the entanglement entropy between the system and the environment. 
Throughout our study we will consider the energy and time scales written in terms of the coupling strength $g$ and its inverse, respectively. In these units, we will consider a band centered at $A=100$ and with width $B=50$ \cite{ogden2008}. Except for zero detunings, the most relevant energy scale of the problem is given by $\Delta$, a quantity that determines the environment memory and, therefore, how Markovian the dynamics is. However, when we have several atoms coupled through the field, the group velocity of light also becomes important, and this quantity is determined by both $\Delta$ and the band width $B$. This will be further discussed in Secs.~\ref{DE} and \ref{inter-atomic}.


\section{Accuracy of the weak-coupling approximation}
\label{WCA}

As already mentioned in the introduction, over the past few years several works have proposed to use atoms interacting with low-dimensional fields to model materials where atomic spin degrees of freedom may couple through the field over long distances \cite{devega2008,douglas2013,tudela2014}. However, most of these studies are based on assuming a weak-coupling approximation to describe the problem. 

The limits of validity of such an approximation are relatively well-known for a single open quantum system or emitter. For instance, for band-gapped spectral densities, non-Markovian effects are important (and therefore the weak-coupling approximation is inaccurate) when the atomic frequencies are in the vicinity of the band-gap edge where the density of states of the e.m. field changes abruptly from zero to a finite value \cite{ruggero2014,devega2014a}. However, less is known about how such performance is affected in the presence of several atoms which are coupled via the e.m. field. In the following we analyze the regimes and situations in which one ought to avoid using a second-order perturbative theory and, therefore, the related (and more restrictive) Markov approximation. 


\subsection{Weak-coupling master equation}
\label{DE}

In order to present the perturbative theory, we consider an initially separable state of the atoms in a state 
$|\psi_\text{S}(0)\rangle$ and the photonic vaccuum $|\textmd{vac}\rangle$:
\bea
|\Psi_\ttot (0)\rangle=|\psi_\text{S}(0)\rangle|\textmd{vac}\rangle.
\label{psitot}
\eea
Then, up to second order in the perturbative parameter, $g$, the evolution of the reduced density matrix of the atoms is given the Master Equation (ME) \cite{breuerbook,rivas2011a,devega2015c}
\begin{align}\frac{d}{dt} \rho^\text{I}_\text{S}(t) = - \int_{0}^{t} \ud s\; \Ttr_\text{B}[H_\text{I}(t),[H_\text{I}(t-s),\rho^I_\text{S}(t) \otimes \rho_\text{B}]],
\end{align} 
where we have considered the interaction picture with respect to $H_0=H_\text{S}+H_\text{B}$ and defined $H_\text{I}(t)=\text{e}^{\text{i}H_0 t}H_\text{I}\text{e}^{-\text{i}H_0t}$.
Replacing the interaction Hamiltonian of our problem (\ref{hEM3}) in such general ME, and going back to the Schr\"odinger picture, we find
\begin{eqnarray} 
\frac{d}{dt} \rho_\text{S}(t) &=&-\text{i}[H_{\eeff}(t),\rho_\text{S}(t)]+g^2\sum_{ij} \gamma_{ij}(t) \bigg[ \sigma_{j} \rho_\text{S}(t)\sigma_{i}^{+}\cr
&-& \sigma_{i}^{+}\sigma_{j}\rho_\text{S}(t)\bigg] + \textmd{h.c.}
\label{MasterEqSystem}
\end{eqnarray}
where $H_{\eeff}(t)= H_\text{S} + H_{\textmd{Lamb}}(t)$ is a sum of the bare system and the Lamb-shift Hamiltonian 
\bea
H_{\text{Lamb}}=g^2\sum_{i,j=1}^N \delta_{ij}(t) \sigma_{i}^{+}\sigma_{j}.
\label{LS}
\eea
In~\eqref{MasterEqSystem} and~\eqref{LS}, we have defined
\begin{align}
\delta_{ij}(t)&=\Iim[\Gamma_{ij}(t)],\\
\gamma_{ij}(t)&=\Rre[\Gamma_{ij}(t)],
\end{align}
with the multi-particle dissipative rate given as 
\bea
\Gamma_{ij}(t)=\int_0^t \ud\tau \;\alpha_{ij}(\tau)\text{e}^{\text{i}\omega_\text{S}\tau},
\label{rates}
\eea
and dependent on the normalized environment correlation functions given by 
\begin{eqnarray} 
&\alpha_{nj}(t) = \sum_{k} \text{e}^{-\text{i} \omega(k)t} \text{e}^{-\text{i} {r}_{nj} k},
\label{Alpha}
\end{eqnarray}
where $r_{nj}=r_n-r_j$. In the continuum limit the correlation functions be written in terms of the density of states 
$D(\omega)=\large|{\partial k}/{\partial \omega(k)}\large|_{k=k(\omega)}$: 
\bea
&\alpha_{nj}(t) =\int_{-\infty}^\infty \ud\omega \;D(\omega)\text{e}^{-\text{i}\omega t}\text{e}^{-\text{i}{r}_{nj} k(\omega)}.
\label{AlphaC}
\eea
With the dispersion relation (\ref{dispersion}), and considering integer inter-atomic distances $r_{nj}=L$, the correlation function can be written as 
\bea
\alpha_{L}(t)\sim \text{i}^LJ_{L}(Bt)\text{e}^{-\text{i}A t}, 
\label{ourcorrel}
\eea
where $J_L(Bt)$ is the Bessel function of order $L$.


The Lindblad or Markovian Master Equation (MME) is obtained by considering~\eqref{MasterEqSystem} and taking the long-time limit of the decaying rates and Lamb shifts, 
$\gamma_{ij}=\gamma_{ij}(\infty)$ and $\delta_{ij}=\delta_{ij}(\infty)$, such that we find 
\begin{eqnarray} \label{MasterEqSystem_Mark}
&&\frac{d}{dt} \rho_\text{S}(t) = -\text{i}  [H_{\eeff},\rho_\text{S}(t)]  \\
&+&g^2\sum_{i,j=1}^N \gamma_{ij}  \left[ \sigma_{i}^{+}\sigma_{j}\rho_\text{S}(t) - \sigma_{j} \rho_\text{S}(t)\sigma_{i}^{+}  \right] + \textmd{h.c.} \nonumber
\end{eqnarray}
The convergence time-scale of the rates to their constant value, and therefore the memory of the environment, depends on the particular problem. To analyze such convergence in our case, we consider the long-time limit expansion of the correlation function~\eqref{ourcorrel}, finding that 
\bea
\alpha_{ij}(t)\sim \frac{\text{e}^{-\text{i}\omega_\text{c} t+\text{i}L\pi/2}}{\sqrt{Bt}}, 
\eea
for $r_i-r_j=L$. Since this function decays polynomially as $(Bt)^{1/2}$, the convergence of the rates~\eqref{rates} to a constant value is in most cases dominated by the oscillations of the integrand produced by the phase $\exp(\text{i}\Delta t)$. Only when $\Delta\approx 0$ the convergence is dominated by the slow polynomial decay, which makes the problem more non-Markovian. 
Moreover, the Markov rate can be approximated as (see details in Appendix \ref{AppB})
\begin{align}
	\Gamma^{\textmd{Mk}}_{ij} \approx \gamma_0 (-1)^{r_{ij}+1} \text{e}^{\text{i}\frac{r_{ij}}{\xi}}, 
	\label{periodicGamma}
\end{align}
where we have defined $\gamma_0=2\pi \xi/B$, with $\xi=h_0\sqrt{B/2\Delta}$. In the Markov limit, 
a negative detuning gives rise to a strict cancellation of the real part of $\Gamma$ and therefore of the dissipation. 
Thus, the dynamics of atoms with frequencies within the gap only depends on a purely imaginary dissipative rate that 
decays exponentially with distance $\xi$ \cite{devega2008}. 
As discussed in the following sections in more detail, a more 
precise account of such dynamics reveals that even for negative $\Delta$ there are some dissipative losses in the 
atomic emission.

\subsection{Effects of multiple excitations}
\label{multiple}

We now analyze a system consisting of having $N=1,2,3$ atoms located at distances $L=1$, by measuring the evolution of the total population in the excited atomic state
\bea
P_\ttot(t)=\sum_{j=1}^N P_j(t),
\label{Ptot}
\eea
with $P_j(t)$ given by~\eqref{population_1}. In addition, we compare the results obtained using the ME with the outcomes of a $t$-DMRG approach in order to check the accuracy of the first equation.


\begin{figure}[ht]
\centering{}\includegraphics[width=1\linewidth]{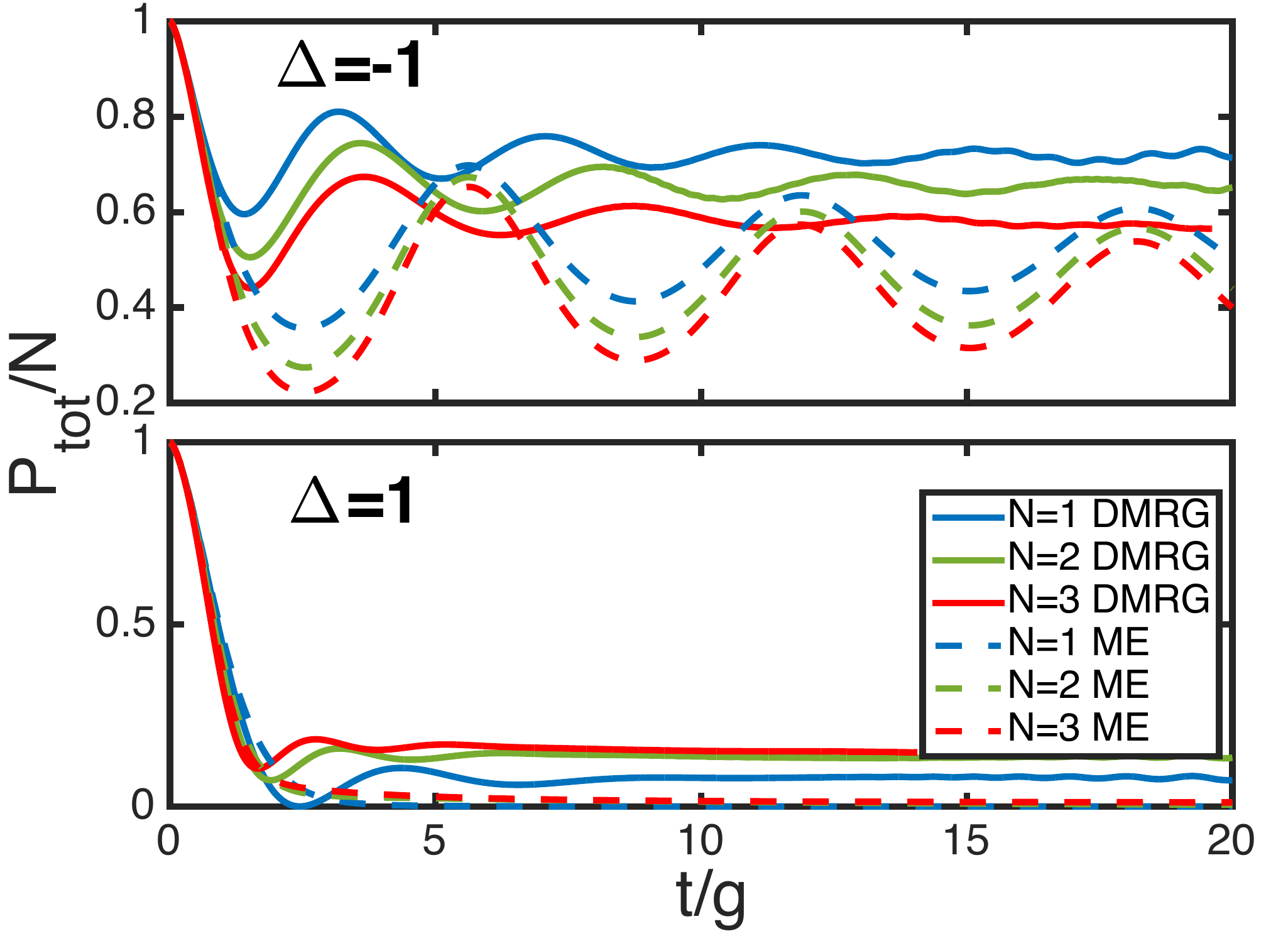}\caption{(Color online) Evolution of the normalized total population a different number $N$ of excited atoms, $P_\text{tot}=\sum_{j=1}^N\langle\sigma_j^+(t)\sigma_j (t)\rangle$, as given by the ME and $t$-DMRG. We consider all the atoms located at distances $L=1$, and the atomic frequencies within the gap ($\Delta=-1$, top panel,  where $N=1,2,3$ correspond to blue (upper), green (middle) and red (lower) lines, respectively) and within the band ($\Delta=1$, bottom panel, where now $N=1,2,3$ corresponds to blue (lower), green (middle) and red (upper) lines, respectively). \label{totalpop}}
\end{figure}

\begin{figure}[!ht]
		\includegraphics[width=1\linewidth]{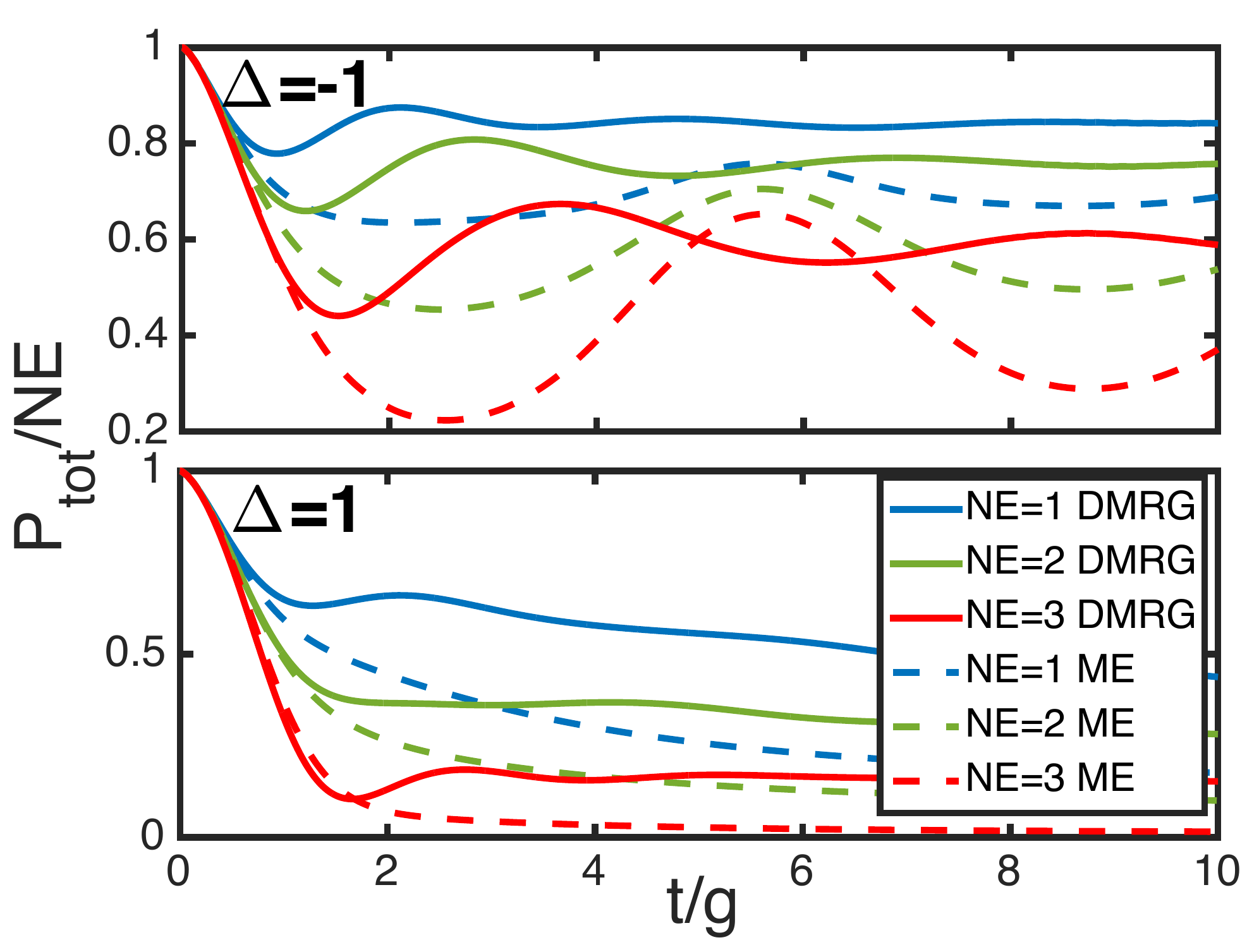}	
\caption{(Color online) Evolution of $N=3$ atoms as given by $t$-DMRG (solid lines) and ME (dashed lines) for the following cases: $|100\rangle$ blue (upper) lines and $|110\rangle$ green (middle) lines, and $|111\rangle$ red (lower) lines. As in Fig.~\ref{totalpop} we consider all the atoms spread at a distance $L=1$.} 
\label{fig:1exVSNex_x1}
\end{figure}

In Fig.~\ref{totalpop} we analyze $N=1,2$ and $3$ atoms with initial states $|\psi_\text{S}(0)\rangle$ where all the atoms are excited, i.e. $|1\rangle$, $|11\rangle$, and $|111\rangle$, respectively.
In contrast, Fig.~\ref{fig:1exVSNex_x1} shows the results for an ensemble of $N=3$ atoms with initial states $|100\rangle$, $|110\rangle$, and $|111\rangle$ having a different number of initial atoms excited, $N_E=1,2,3$. In both figures we observe that for atomic frequencies in the gap ($\Delta=-1$) and to a 
lesser extent in the band ($\Delta=1$) there is no complete dissipation of the total population \footnote{For $\Delta=0$ (not shown here) we find similar dynamics as for $\Delta=-1$, just with a slower dynamics. As already mentioned the reason is the downward shift of the system frequency. This is true for all the other cases analyzed in the following and for this reason we will not comment on it anymore.}. This is because in the gap and in the band area near the edge, the density of states varies very abruptly, giving rise to strong non-Markovian effects in the system dynamics \cite{navarrete2010,devega2014a}. Because of this, 
a fraction of the emission is radiated in the form of non-propagating modes that remain exponentially localized near the atoms within a region determined by the localization length $\xi$. These localized excitations are continuously reabsorbed and re-emitted by the atom, giving rise to what is known in the literature as photon-atom bound state \cite{pcbook,quang1997,florescu2001,devega2005,tong2010,sanchez2017},  which has been also analyzed in surface plasmon polaritons \cite{yang2017}. In particular for a single atom, the stationary state -- once the propagating photons are irreversibly dissipated -- would correspond to the atom-photon bound state, which can be written as \cite{woldeyohannes2003,tong2010}
\bea
|\Psi_{t\rightarrow\infty}\rangle=c_1|1,\textmd{vac}\rangle+\sum_k b_k|0,1_k\rangle,\; b_k=\frac{g c_1}{e_b-\omega_k}. 
\label{single_excit}
\eea
where $e_b$ is the atom-photon bound state energy, which satisfies the usual equation for a localised state level in a continuum (see, e.g., \cite{mahan1981})
\bea
e_b-\omega_s=g^2\sum_k\frac{1}{e_b-\omega_k}.
\label{bounden}
\eea
Decreasing $|\Delta|$ increases the photonic part of the superposition and its localization length $\xi$ (see also the discussion in \cite{navarrete2010} for a closely related system). On the other hand the more localized the emission is, the more population is preserved in the atom in the long time limit \cite{woldeyohannes2003,devega2014a}. We can also anticipate that the longer the localization length the more atoms will be connected through the field, as described by the rates~\eqref{periodicGamma} in the Markov case, and therefore the stronger the collective or cooperative effects will be.

We now address the question of the system robustness against dissipation when adding more atoms o more excitations. We observe that in general the presence of ground state atoms slightly inhibit dissipation (compare Figs.~\ref{totalpop} and \ref{fig:1exVSNex_x1}).
On the other hand, as expected and as shown in Fig. \ref{fig:1exVSNex_x1} (see also Fig.~\ref{totalpop}) the dissipation increases, by increasing the number of excited atoms.

Concerning the question addressed in the present Section, the most important feature of the presence of a atom-photon bound state is the failure at \textit{long times} of the ME. Indeed we see that the latter is able to give very accurate results only on the short time scale $t\le g^{-1}$. For longer times the ME underestimates the localization of light both for frequencies in the band and in the gap, predicting a much smaller atomic population than the one given by the exact result.

\subsection{Effects of the inter-atomic distance}	
\label{inter-atomic}	

To study the accuracy of the ME for different inter-atomic distances we consider the simplest case of two atoms initially in a product state 
$|\psi_\text{S}(0)\rangle=|10\rangle_{L}$, where the first atom is in the excited state 
$|1\rangle$ and the second atom, located at a varying distance $L$, is in the ground state $|0\rangle$. Other interesting cases for population and entanglement preservation the long times dynamics are considered in the next Section.
Given the number conserving nature of the Hamiltonian the total wave function can be written as
\begin{align}
|\Psi_t\rangle=(c_1|10\rangle+c_2|01\rangle)\otimes|\textmd{vac}\rangle+\sum_k b_k |001_k\rangle
\label{2atomstate}
\end{align}
and the problem can be conveniently treated with ED. As discussed further in Appendix \ref{AppBound}, for atomic energies within the gap and in the the band-edge region some of the emitted radiation will localice nearby the atoms and the steady state will be written in terms of  bound states.
We then compute the population of the excited state of atom $j$ as  
\bea
P_j(t)=\langle\sigma_j^+(t)\sigma_j(t)\rangle, 
\label{population_1}
\eea
by considering the ME and the ED results, obtaining $P_{j}^{\ME}$ and $P_{j}^{\NE}$, respectively. The error produced by the weak coupling approximation is then quantified as the time averaged difference between these two quantities, i.e.   
\bea
E(L, \omega_\text{S}) &=& \frac{1}{T} \int_{0}^{T} \ud t 
\left| P^{\ME}_{1}(t,L, \omega_\text{S}) - P^{\NE}_{1}(t,L, \omega_\text{S}) \right|.\cr
&&
\label{errorMEE}
\eea

\begin{figure}[ht]
\centering
\includegraphics[width=1\linewidth]{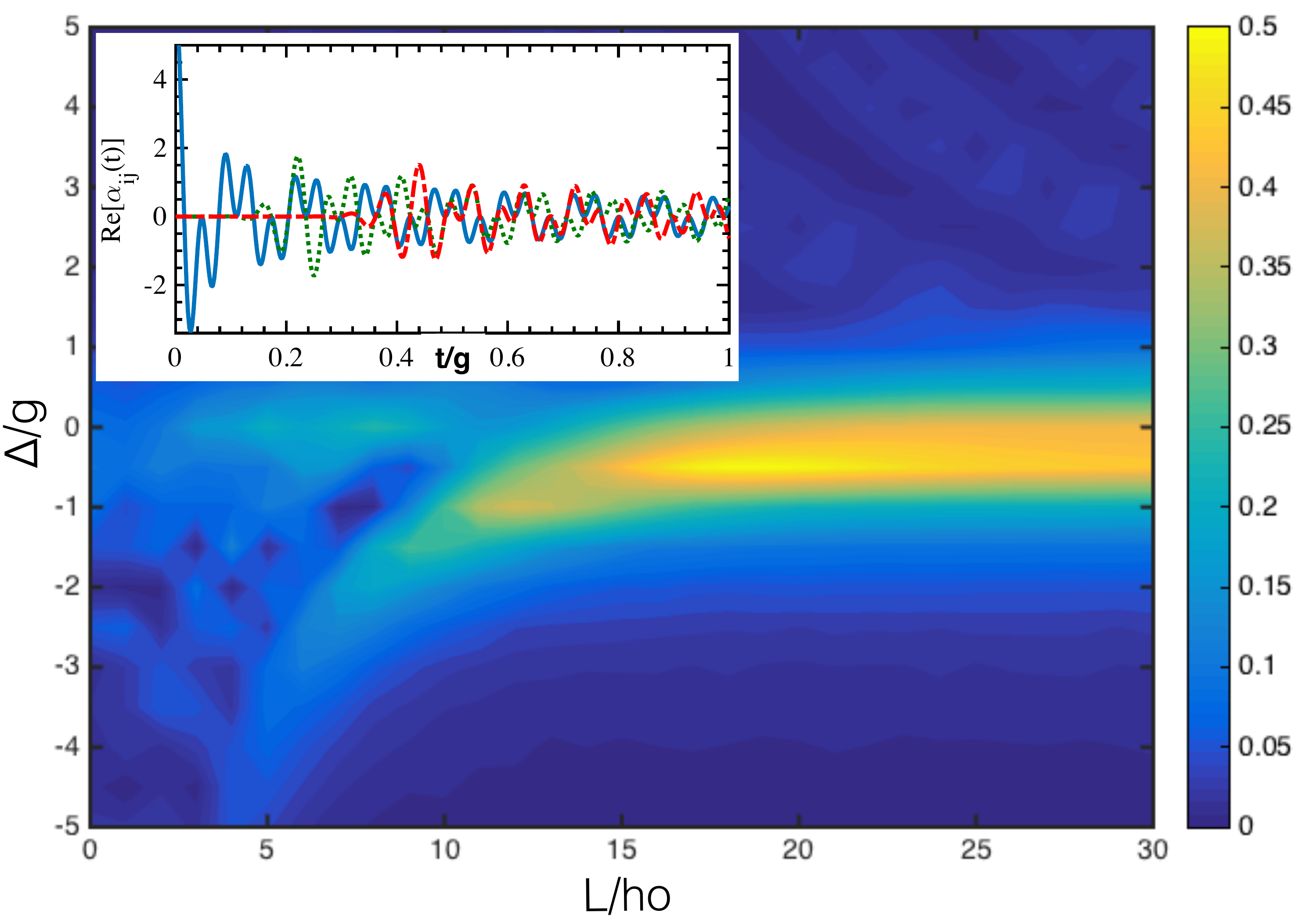}\caption{(Color online) Contour plot representing the difference between the population of atom $1$ (which is initially excited) as given by the ME and a numerically exact solution based on exact diagonalization and quantified by $E(L,\omega_\text{S})$ given by (\ref{errorMEE}). The $y$-axis represents the frequency shift between the atomic and gap-edge frequencies, $\Delta=\omega_\text{S}-\omega_\text{c}$, where $\omega_\text{c}=A-B$. The inset represents the real part of the correlation function (\ref{Alpha}) for increasing distances $r_{i}-r_{j}=0,10,20$ in solid blue, dotted green, and dot-dashed red respectively. \label{error}}
\end{figure}
The contour plot of such quantity is displayed in Fig.~\ref{error} for different values of the atomic frequency $\omega_\text{S}$ and the 
inter-atomic distance $L$.
The time $T$ in~\eqref{errorMEE} is chosen such that for the corresponding parameters the system can be considered to have reached a steady state. In those regimes where oscillations are present in the long time limit such that the convergence to the steady state is very slow (like when the atomic frequencies are within the gap) a time average of the curve is further considered. 

The contour shows that the ME is a very good estimation for relatively short distances and atomic frequencies $\omega_\text{S}$ that are far detuned from the gap-edge frequency $\omega_\text{c}$. 
However, it can also be noticed that the maximum error does not occur at $\Delta=0$ but displays a more subtle structure. This can be analyzed by distinguishing two different limits, large inter-atomic distances ($L>10$) and short inter-atomic distances ($L<10$). 

\subsubsection{Large inter-atomic distances}

At large distances the failure of the ME predominantly occurs at the band-gap edge. However, as discussed previously, the system energy is slightly corrected by the dressing of the environment, and therefore the resonance to the band-gap edge is shifted to values of $\Delta$ slightly below zero. The energy corrections are very similar to the ones found in  \cite{diaz2015} for atoms in free space while the single atom correction is also analyzed in the similar model of Ref.~\onlinecite{navarrete2010}.

Fig.~\ref{error} shows that the second order ME fails more for atomic frequencies at the band edge and for atoms located at larger distances. The bigger failure of the ME at larger distances is similar to the result discussed in \cite{zheng2013,guimond2016} for the Markov approximation. 
Such a failure is linked to the fact that the Markov approximation does not describe correctly the 
presence of finite time delays in the atom-atom interactions mediated by the field, 
due to the relatively small speed of light within the CROW. 
Indeed, as shown in the inset of Fig.~\ref{error}, the correlation function $\alpha_{nj}(t)$ 
is zero before a certain time delay $t^{nj}_\text{d}$ that is found to grow with the distance 
$r_{nj}=r_n-r_j$.  

To better illustrate the effect of the delay in atom-atom interactions in the validity of 
the weak-coupling approximation, we consider the time-evolution equation of the two-point correlation functions $\sigma_n^+\sigma_l$ in the interaction picture with respect to the environment $H_\text{B}$. In this picture, the equation of motion for a system operator $A$ is $
\frac{d}{dt}A(t) = \text{i}{\mathcal U}^\dagger (t)\left[ H, A \right]{\mathcal U}(t)$,
with ${\mathcal U}(t)= \text{e}^{-\text{i}H t}\text{e}^{\text{i}H_\text{B} t}$. For the system correlation functions we find \cite{devega2015c}%
\begin{align}
&\frac{d}{dt}\langle \sigma_n^+(t)\sigma_l(t) \rangle=-g\langle \hat{v}_n^{\dagger} (t)\sigma^z_n(t)\sigma_{l}(t) \rangle
-g\langle \sigma_{n}^{+}(t)\sigma^z_l(t)\hat{v}_l(t) \rangle\cr
-&g^2 \sum_{j=1}^N \int_{0}^{t} \ud \tau \;\alpha_{nj}^{*} (t- \tau) \langle\sigma_{j}^{+}(\tau)\sigma^z_n(t) \sigma_{l}(t) \rangle \nonumber \\
-&g^2\sum_{j=1}^N \int_{0}^{t} \ud \tau \;\alpha_{jn} (t- \tau) \langle \sigma_{n}^{+}(t) \sigma_l^z(t)\sigma_{j}(\tau)\rangle,
\label{final}
\end{align}
where we have introduced the noise operator $\hat{v}_j(t) = \text{i}  \sum_{k} \text{e}^{\text{i} kr_{j}} \text{e}^{-\text{i}  \omega(k) t} a_{k}(0)$. Moreover, we note that in~\eqref{final}, the averages are performed with respect to the initial state. Considering the initial condition (\ref{psitot}) the first two terms in Eq. (\ref{final}) vanish 
		\bea
		\langle \hat{v}_n^{\dagger} (t)\sigma^z_n(t)\sigma_{l}(t) \rangle=\langle \sigma_{n}^{+}(t)\sigma^z_l(t)\hat{v}_l(t) \rangle=0.
		\label{vanish}
		\eea 
Despite such simplification, the exact Heisenberg equation~\eqref{final} is not closed since it depends on two-time correlation 
functions that should be computed independently. The evolution of two-time correlations is in turn coupled to the 
evolution of three-time correlations and so on, which makes the full problem not tractable without considering any 
approximation. 
One of the most common approximations is that of weak coupling, which treats the system-environment coupling $g$ as a 
perturbative parameter to perform expansions. 
In our case, we will expand up to second order. Since the two-time correlations in~\eqref{final} are 
already in a term of at least order $g^2$, we can approximate
\bea
&&\langle\sigma_{j}^{+}(\tau) \sigma_n^z(t)\sigma_{l}(t) \rangle=\langle {\mathcal U}^{-1}(t){\mathcal U}^{-1}(\tau-t)\sigma_j^+{\mathcal U}(\tau-t)\cr
&\times&\sigma^z_n\sigma_l{\mathcal U}(t)\rangle\approx\langle {\mathcal U}^{-1}(t)V_{\tau-t}\sigma_j^+\sigma^z_n\sigma_l{\mathcal U}(t)\rangle+{\mathcal O}(g^2).\cr
&&
\label{WC}
\eea
Here, we have defined the operator $V_t A=\text{e}^{\text{i}H_\text{S} t}A\text{e}^{-\text{i}H_\text{S} t}$, which acts only on the first system operator next to it, $A$.
Considering this in~\eqref{final}, we obtain the following Heisenberg 
equation up to second order in $g$
		\begin{align}
&			\frac{d}{dt}\langle \sigma_n^+(t)\sigma_l(t) \rangle=\cr
&- g^2\sum_{j=1}^N \int_{0}^{t} \ud \tau\; \alpha_{nj} (t- \tau) \text{e}^{\text{i}\omega_j(\tau-t)} \langle\sigma_{j}^{+}(t)\sigma^z_n(t) \sigma_{l}(t) \rangle \nonumber \\
			&- g^2\sum_{j=1}^N \int_{0}^{t} \ud \tau\; \alpha_{jn} (t- \tau) \text{e}^{-\text{i}\omega_j(\tau-t)} \langle \sigma_{n}^{+}(t) \sigma_l^z(t)\sigma_{j}(t)\rangle,
			\label{final_2}
		\end{align}
which depends on normalized correlation functions and therefore display explicitly the order in $g$ of each term.  

Let us now consider for simplicity a single atom coupled to a $\delta$-correlated 
environment with $\alpha_{nj}(\tau)=g^2\gamma\delta(\tau)$ 
(here we neglect the principal value part for simplicity in the explanation). 
The weak-coupling approximation is very accurate in this case: when replacing such a $\delta$-correlation in the 
exact equation~\eqref{final}, we arrive at the same result as when the replacement is made in 
the second-order~\eqref{final_2}. If we now consider several atoms separated by a 
finite distance, the weak-coupling approximation gets inaccurate even for $\delta$-correlated 
environments, in which case one can write~\eqref{AlphaC} as 
(see Appendix \ref{delta_correlated} for details) 
\bea
\alpha_{nj}^{*} (\tau)= \delta(\tau-t^{nj}_\text{d}),
\label{delta_correlated_correlation}
\eea 
with $t_\text{d}^{nj}=r_{nj}/v_\text{g}$ and $v_\text{g}$ the  group velocity of light in the medium.

Replacing this in the first integral term of the exact equation (\ref{final}), for instance, we find
\bea
g^2\sum_{j=1}^N& \int_{0}^{t} \ud \tau\; \alpha_{nj}^{*} (\tau) \langle\sigma_{j}^{+}(t-\tau)\sigma^z_n(t) \sigma_{l}(t) \rangle\cr
&=g^2\sum_{j=1}^N \langle \sigma_j^+(t-t^{nj}_\text{d})\sigma^z_n(t)\sigma_l(t)\rangle.
\eea
Hence, the equation still depends on two-time correlation functions. However, if we consider the same term within the weak-coupling equation~\eqref{final_2}, we find 
\bea
g^2\sum_{j=1}^N& \int_{0}^{t} \ud \tau \alpha_{nj}^{*}(\tau)\text{e}^{-\text{i}\omega_\text{S}\tau} \langle\sigma_{j}^{+}(t)\sigma^z_n(t) \sigma_{l}(t) \rangle\cr
&\approx  g^2\sum_{j=1}^N \text{e}^{-\text{i}\omega_\text{S} t^{nj}_\text{d}}\langle \sigma_j^+(t)\sigma^z_n(t)\sigma_l(t)\rangle.
\eea
Hence, in the $\delta$-correlated case, the weak-coupling approximation corresponds to the assumption that 
\begin{align}\nonumber
&\langle\sigma_j^\dagger(t-t^{nj}_\text{d})\sigma^z_n(t)\sigma_l(t)\rangle\\
\approx&\;\text{e}^{-\text{i}\omega_\text{S} t_\text{d}^{nj}}\langle\sigma_j^+(t)\sigma^z_n(t)\sigma_l(t)\rangle,
\end{align}
which is clearly not accurate for large time delays 
as compared to the evolution time $T_S$. 
We shall note that the retardation effects are only significant for light within materials giving rise to a slow group velocity, and therefore to large retardation times. In our case the group velocity is frequency dependent, $v_\text{g}(\omega)=d\omega_k/dk\sim Bh_0\sqrt{2\omega_\text{S}\Delta(\omega)}$, where $\Delta(\omega)=\omega-\omega_\text{c}$ and $\omega$ is the frequency of light. For the resonantly emitted photons, i.e. those with $\omega=\omega_\text{S}$, we find that the smaller the band width or the smaller the  detuning $\Delta(\omega_\text{S})=\Delta$, the slower the group velocity. In contrast, in vacuum, when $v_\text{g}=c$ the retardation effects can in general be neglected \cite{lehmberg1970}.

\subsubsection{Small inter-atomic distances}

More subtle is the situation for smaller inter-atomic distances. In general, the ME appears to be more accurate than for long distances. However, Fig.~\ref{error} shows that the values of $L$ and $\Delta$ at which the second-order approximation fails more appear in two branches. The first branch corresponds again to approximately the band edge, but there is a second branch line that curves towards smaller values of $\Delta$. Such a line is approximately given by the maximal inter-atomic distance in which two atoms are connected for a given $\Delta$, $L\approx  2|\xi|$. Its existence 
is again related to the presence of stationary localised atom-photon bound states
as we will detail in the next Section. For larger $L>2|\xi|$ there are no atom-atom interactions, and therefore no retardation effects, and thus since additionally we are far from the band-gap edge the ME is rather accurate. For smaller $L$ the atoms interact with one another, but the distances are not large enough so as to produce relevant retardation effects, and thus the ME still remains accurate. Hence, the line $L\approx 2|\xi|$ signals the limiting condition at which both the atom-atom interactions and retardation effects are important. Also, such a condition also signals the region in which collective or cooperative effects are relevant. This is confirmed by Fig.~\ref{fig:dicke} in App.~\ref{Dicke_sec}, which shows that the accuracy of the Dicke approximation (which roughly speaking always considers the presence of collective effects) is also delimited precisely by this line.

\section{Entanglement and population dynamics}
\label{variation}

In the previous section we have shown the importance of analyzing our system beyond the weak-coupling approximation, particularly within the band-gap edge region and at long atomic distances. In the following, using $t$-DMRG, we 
study the dynamics of the 2-level systems, focusing on the population and the entanglement evolution and their dependence on the distance.

To this aim, we consider $N=2$ atoms having resonant frequencies $\omega_\text{S}$ either within the gap or within the band. The initial state is taken of the form (\ref{psitot}), with $|\psi_\text{S}(0)\rangle$ given by  
	\begin{itemize}
			\item $|\psi_{1}\rangle= \ket{11}$, corresponding to both spins excited,
			\item $|\psi_{2}\rangle = \frac{1}{\sqrt{2}} \left( \ket{10} + \ket{01} \right)$, corresponding to a maximally entangled state,
			\item $|\psi_{3}\rangle = \frac{1}{2} \left( \ket{0}_{1} + \ket{1}_{1} \right) \otimes \left( \ket{0}_{2} + \ket{1}_{2} \right)$, with each spin in a superposition of the two basis states.
		\end{itemize}
We study the dynamics of the system for different positions $r_2=L$ of the second atom. 

\begin{figure}[!ht]
		\includegraphics[width=1.\linewidth]{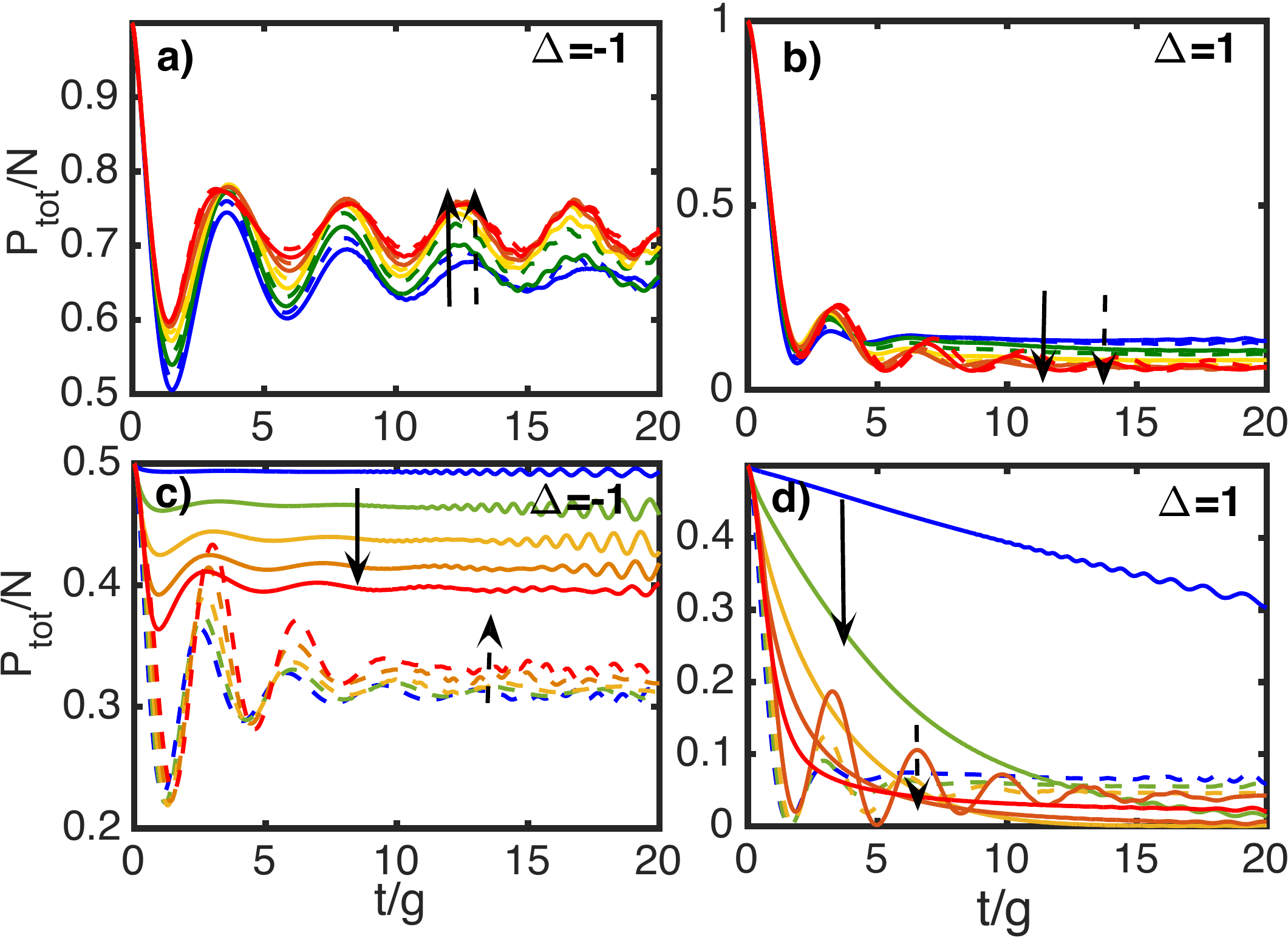}	
\caption{(Color online) Evolution of the population for the initial condition $|\psi_1\rangle$ (two upper panels) and $|\psi_2\rangle$ (two lower panels). The curves, corresponding to  different inter-atomic separations $L$, are displayed in a rainbow scale that goes from blue ($L=1$) to red ($L=10$), also indicated by the arrows. We also distinguish between even separations (solid lines) and odd separations (dashed lines). The presence of an even-odd difference for the initially entangled state is related to sub and superradiant states as explained in the main text.} 
\label{fig:pop_x_is1}
\end{figure}

Fig.~\ref{fig:pop_x_is1}(a,b) displays the dynamics of the total atomic population~\eqref{Ptot} considering an initial condition $|\psi_1\rangle$  
\footnote{Notice that we analyse also the case $\Delta=0$, and it looks pretty much like in the gap, but with a much slower dynamics.}.
Aside from the trivial fact that the dissipation is stronger in the band (for $\Delta=1$), than in the gap ($\Delta=0-1$), the new aspect is that the inter-atomic distance plays a different role in the two (band or gap) cases: atoms within the band dissipate less when they are close to each other than further away, while for atoms within the gap the opposite is observed. The reason 
is that if $\omega_\text{S}$ is in the band, the environment shifts the frequency further away from the band edge when $L$ diminishes, which gives rise to less localization and dissipation. When $\omega_\text{S}$ is in the gap, the environment pulls the frequency closer to the edge when $L$ diminishes, giving rise to a larger localization length $\xi$ and to more dissipation.

In Fig.~\ref{fig:pop_x_is1}(c,d) we analyze the same quantity but considering an initial entangled state $|\psi_2\rangle$.  
In this case, we observe that the steady state changes periodically with the distance in such a way that for odd distances the steady state population is always larger than that for even ones. We shall emphasize that this behavior is mainly present when the initial state is either $\ket{\psi_{2}}$ or $\ket{\psi_{3}}$ (discussed in App.~\ref{AppD}), while it is much more difficult to discern for the initial state $\ket{\psi_{1}}$ displayed in Fig.~\ref{fig:pop_x_is1}. In other words, it happens 
only when the initial state contains coherences. 
The difference in the decaying for odd and even distances observed in Fig.~\ref{fig:pop_x_is1}(c,d) can be analyzed by considering the evolution of the atomic coefficients in Eq. (\ref{2atomstate}) (see also discussion in App.~\ref{AppBound}).
Indeed, because of the symmetry of the Hamiltonian the exact problem decouples into two variables $c_{\pm}=(c_1\pm c_2)/\sqrt{2}$ \cite{diaz2015}, with evolution equations
\bea
i\partial_t c_{\pm}=\omega_\text{S} c_{\pm}-\text{i}\int_0^t \ud s\; \alpha_{\pm}(t-s)c_{\pm}(s),
\label{pm}
\eea
where we have defined $\alpha_{\pm}(t)=2\alpha_{11}(t)\pm 2\alpha_{12}(t)$, and the correlation functions $\alpha_{11}(t)\equiv \alpha_{0}(t)$ and $\alpha_{12}(t)\equiv \alpha_{L}(t)$ depend only on the atomic distance $L$ and are given by Eq.~\eqref{ourcorrel}. Formally, the above system can be solved by considering that its Laplace transform is just
\bea
[s+\text{i}\omega_\text{S}+ \alpha_{\pm}(s)]c_{\pm}(s)=c_{\pm}(0),
\eea
where $\alpha_{\pm}(s)=2\alpha_{0}(s)+2\alpha_{L}(s)$ is the Laplace transform of $\alpha_{\pm}(t)$, and $\alpha_{L}(t)$ is given by~\eqref{ourcorrel}. 
In particular the zeros of function $s+\text{i}\omega_\text{S}+ \alpha_{\pm}(s)$ 
gives  system energies and decaying rates.
Obviously when $L=0$, $\alpha_{12}(t)=\alpha_{11}(t)$, one has that the coefficient $c_{-}(t)$ only oscillates in time with the bare frequency $\omega_\text{S}$ and without any decay. Therefore starting with the appropriate entangled atomic state one can obtain no decay of the atomic population.

\begin{figure}[!ht]
		\includegraphics[width=1\linewidth]{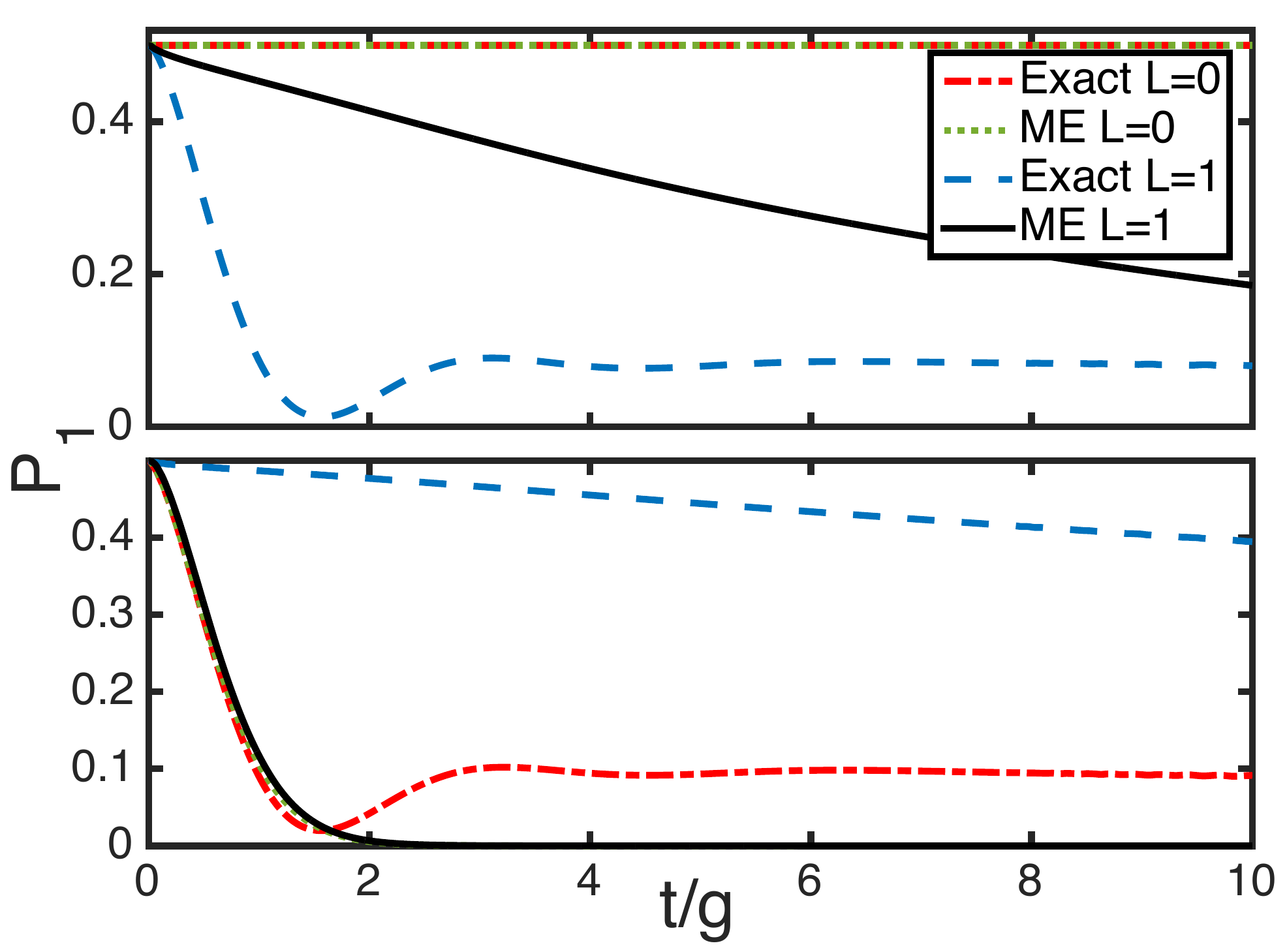}	
\caption{(Color online) Evolution of the population for $\Delta=1$ considering the initial condition $|\tilde{\psi}_2\rangle$ (upper panel) and $|\psi_2\rangle$ (lower panel) and two different atomic distances, $L=0$ and $L=1$, and as predicted by the exact result and the approximated ME. }
\label{sub-superradiance}
\end{figure}

To see this, we represent in Fig.~\ref{sub-superradiance} the evolution for $c_{-}(0)=1$, i.e.~corresponding to initial state $|\tilde{\psi}_2\rangle=1/\sqrt{2}(|10\rangle- \ket{01})$ (upper panel), and for $c_{+}(0)=1$, corresponding to initial state $|\psi_2\rangle$ (lower panel). 
The results confirm that for $L=0$ the initial state $|\tilde{\psi}_2\rangle$ is subradiant (actually it does not decay) while $|\psi_2\rangle$ is superradiant. 
Turning now to the case $L=1$, we shall realize that for such a distance and immediately after the system is coupled to the environment the states we have are 
\bea
&&|\tilde{\psi}_2\rangle\rightarrow 1/\sqrt{2}(\ket{10} - \text{e}^{\text{i}\pi L} \ket{01})= |\psi_2\rangle,\cr
&&|\psi_2\rangle\rightarrow 1/\sqrt{2}(\ket{10} + \text{e}^{\text{i}\pi L} \ket{01})= |\tilde{\psi}_2\rangle.
\eea 
In other words, $|\tilde{\psi}_2\rangle$ becomes superradiant and $|\psi_2\rangle$ becomes subradiant or a dark state. 
We find that the ME predicts the same subradiant curve as the exact method for $L=0$ (not distinguishable curve), while it underestimates the superradiance occurring for $L=1$. Interestingly, in the lower panel the ME gives a good estimation of the superradiance for $L=0$ and an extremely poor description of the subradiance for $L=1$. Hence, subradiance appears to be due to an interference effect produced by the presence of system-environment entanglement, which is not well-accounted for by the Born approximation included in the weak-coupling ME. A further analysis of the periodic behavior observed in Fig.~\ref{fig:pop_x_is1}(c,d) is provided in Appendix~\ref{AppC}.

\begin{figure}
		\includegraphics[width=1\linewidth]{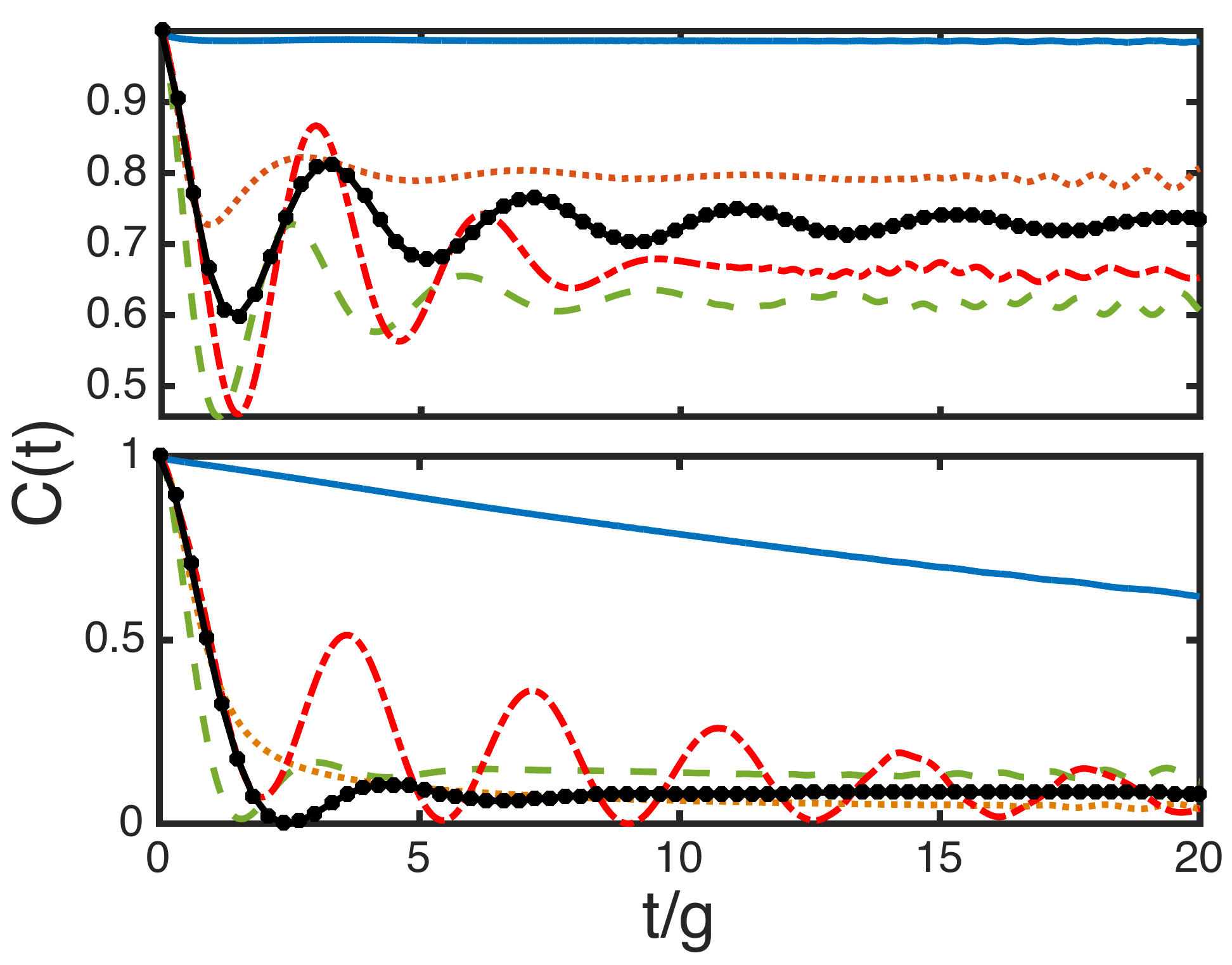}	
\caption{(Color online) Evolution of the concurrence starting from an initially entangled state $|\psi_2\rangle$. The upper and lower panel corresponds to $\Delta=-1$ and $\Delta=1$ respectively. In the same way as in Fig.~\ref{fig:pop_x_is1}(c,d), the solid blue, dashed green, dotted orange, and dot-dashed red correspond to inter-atomic distances $L=1,2,9$ and $10$ respectively. In addition, the black squared curve represents the concurrence of atoms coupled to independent environments as given by~\eqref{independent}.}
\label{concurrence}
\end{figure}

Naively one could think that the evolution of the population does not provide any information concerning how the atomic entanglement may evolve. However, as we discussed above, a non-vanishing steady excited state population requires the atoms to be in an entangled state, and eventually, as discussed in the previous section, to the presence of an atom-photon bound state. This feature has been previously analyzed for the case of atoms equally coupled to a common environment (i.e. the Dicke limit) \cite{tong2010} and for atoms coupled to independent reservoirs \cite{bellomo2009}.

Here we are interested in studying the reduction of the entanglement due to the coupling with the bath.
We consider again the initial atomic state $|\psi_2\rangle$ and study the evolution of the concurrence for different value  
values of the atomic separation.
For a mixed state of two atoms given by the reduced density matrix $\rho_s$, the concurrence is defined as 
\bea
C(t)={\textmd{max}}(0,\lambda_1-\lambda_2-\lambda_3-\lambda_4),
\eea
in which $\lambda_1,\cdots,\lambda_4$ are the eigenstates in decreasing order of the Hermitian matrix $R=\sqrt{\sqrt{\rho_s}\tilde{\rho}_s\sqrt{\rho_s}}$, with $\tilde{\rho}_s=(\sigma_y\otimes\sigma_y)\rho_s^*(\sigma_y\otimes\sigma_y)$ the spin-flipped state of $\rho_s$
The results are reported in Fig. \ref{concurrence}. In the same figure, in order to understand better the role of collective effects, we also 
present the results when the two spins are coupled to independent reservoirs. 
For such a case of independent reservoirs, the concurrence can be exactly computed as \cite{yu2007,bellomo2009},
\bea
C(t)=2{\textmd{max}}\{0,K_1(t),K_2(t)\},
\label{independent}
\eea
where we have defined 
\bea
K_1(t)&=&|q(t)|^2\{\rho_{23}(0)-\sqrt{\rho_{11}(0)}\cr
&\times&[\rho_{44}(0)+\rho_{11}(0)|u(t)|^2\cr
&+&(\rho_{22}(0)+\rho_{33}(0))|u(t)|^2]^{1/2}\},\cr
K_2(t)&=&|q(t)|^2\{\rho_{14}(0)-\sqrt{\rho_{22}(0)+\rho_{11}(0)}\cr
&\times&[\rho_{33}(0)+\rho_{11}(0)|u(t)|^2]^{1/2}\},
\eea
Here, we have defined the matrix elements of the reduced density matrix in the computational basis for the two atoms, i.e. ${\mathcal B}=\{|1\rangle\equiv|11\rangle,|2\rangle\equiv|10\rangle,|3\rangle\equiv|01\rangle,|4\rangle\equiv|00\rangle\}$. Furthermore, we have defined $u(t)=1-|q(t)|^2$, where $q(t)$ obeys the differential equation
\bea
\frac{dq(t)}{dt}=-g^2\int_0^t \ud s\; \alpha(t-s)\text{e}^{\text{i}\omega_\text{S}(t-s)}q(s)
\eea
As noted above, long time limit entanglement is fully linked to the presence of a long time limit population in the excited state, which for a single atom is simply $|q(t)|^2$. In addition, as discussed by Tong et al \cite{tong2010} for the opposite limit in which both atoms are coupled to a common reservoir with the same coupling strength (i.e. the Dicke limit) the concurrence is also proportional to $|q(t)|^2$. The same occurs for  the intermediate regimes described in our case, i.e. atoms inhomogeneously coupled to a common reservoir, as it can be seen by comparing the entanglement dynamics of Fig. 7 with the populations for the same initial state described with Fig.\ref{fig:pop_x_is1}. Thus, the persistence of entanglement in the long time limit is linked to the presence of a photon-atom bound state \cite{tong2010,yang2017}.
For that reason, the concurrence is preserved for gap frequencies in general. Moreover, as expected the subradiant states, due to their dark state nature preserve the entanglement even for in band atomic frequencies. In this regard, just as in the case of the population, when atoms are separated by an odd spacing the steady state entanglement is higher (for frequencies in the gap) and the decay is slower (for frequencies within the band). Interestingly, when comparing the dynamics to that of the case of independent reservoirs one finds that only at odd distances the presence of collective effects is beneficial for further preserving the entanglement.

\begin{figure}
		\includegraphics[width=1\linewidth]{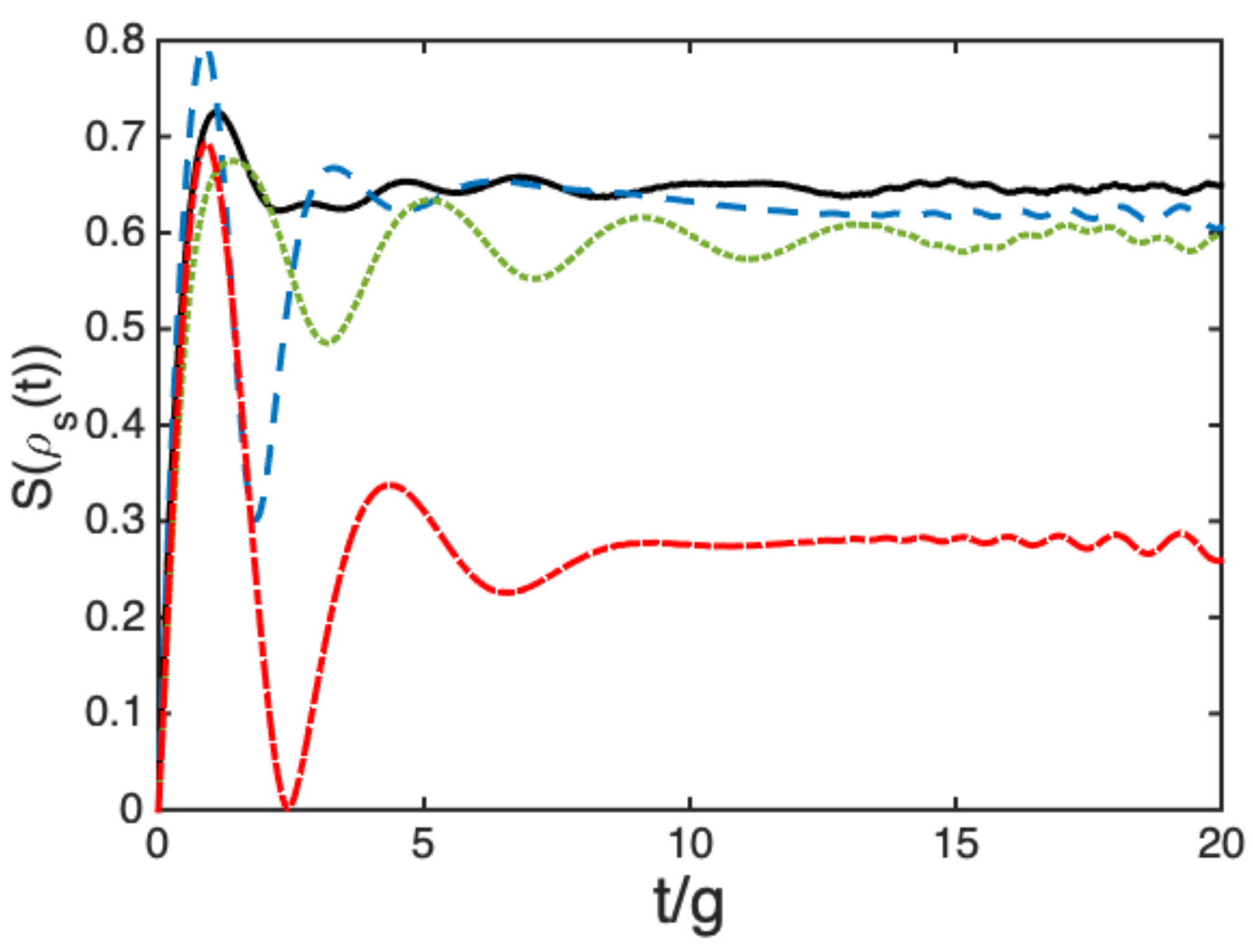}	
\caption{(Color online) Evolution of the von-Neumann entropy, which accounts for the degree of entanglement between the atomic system and the environment. The different curves correspond to: $N=1$ atom with $\Delta=1$ (red dot-dashed curve) and $\Delta=-1$ (green dotted); and $N=2$ atoms ($L=1$) with $\Delta=1$ (blue dashed) and $\Delta=-1$ (solid black).}
\label{VN}
\end{figure}

Finally, we analyze in Fig.~\ref{VN} the entanglement between the atoms and the environment, as measured by the von-Neumann entropy. We consider  $N=1,2$ atoms with initial state $|1\rangle$ and $|11\rangle$ respectively. Interestingly, even for frequencies within the band the system-environment entanglement grows to a non-vanishing steady state, although this is smaller than in the gap case. In addition, such entanglement is in general larger for a larger number of atoms. 

\section{Conclusions}
\label{conclusions}
We have performed a complete analysis of the dynamics of a set of atoms interacting with a 1D electromagnetic field which has  a band gap dispersion. In detail, we discuss the following threeIn this case, t main aspects.
\begin{itemize}
\item \textit{Performance of the weak coupling approximation:} We analyze the performance of the  commonly used weak-coupling approximation by considering a weak-coupling ME and comparing its results with numerically exact methods, $t$-DMRG and ED. 
For a single emitter it is well known that the ME fails in describing the system's dynamics close to the band-gap edge. Our analysis generalise such a result to many atoms where not only their number, but also their distance is varied. We have shown that aside from the trivial case of very large detuning from the band edges and large interatomic distances, the ME is unable to properly take into account the long time dynamics of the multi-emitter system. The reason is related to the impossibility of the ME approach to take into account retardation effects when atoms are connected through the field, and also the presence of atom-photon localised bound states, which are at the origin of the long time finite population.


\item \textit{Robustness against dissipation and decoherence:} We have studied the system population and entanglement dynamics when considering different number of atoms, initial conditions and atomic separations. In all situations we have confirmed previous studies for the single atom case by showing that the atomic population is more preserved in the long time limit for atomic frequencies in the gap. As a new result we have observed that, for a fixed number of initial excitations, there is less dissipation the more atoms we have. However, when fixing the number of atoms we have seen that there is more dissipation the more excitations are initially in the system. Moreover, we find that the system entanglement is more preserved when atoms are coupled to a common environment but only when separated at odd distances. 

\item \textit{Periodicities in the relaxation rates:} Thus, whether atoms are separated even or odd distances is relevant for both the steady state and the dynamics. Indeed, the periodicity in the dispersion relation gives rise to periodicities with the distance in the field-mediated atom-atom interactions, and those in turn give rise to a periodicity in the system relaxation rates. 
Such periodicity in the system relaxation rates has been previously described, but with approximated methods and without taking into account the effect of having different initial conditions. Here, with $t$-DMRG we have shown that relaxation rate periodicities are only significant for initial states already containing coherences, since this makes the effect of the periodic atom-atom interaction between different atoms more significant. 
\end{itemize}
In summary, this work provides what we consider is the first analysis of the dynamics of atoms coupled to 1D fields that on the one hand explores the effect of having different number of atoms and initial conditions, and on the other hand analyzes the accuracy of two of the most commonly used approaches: the weak-coupling and the Dicke approximations. Regarding the last, we unveil the link existing between the validity of the weak-coupling approximation and the presence of collective effects. 

\begin{acknowledgements}
We are very grateful to N.O. Linden for interesting discussions during the development of this work. I.D.V. was financially supported by the Nanosystems Initiative Munich (NIM) (project No. 862050-2) and  the DFG-grant GZ: VE 993/1-1.
\end{acknowledgements}

\appendix

\section{Obtaining the ladder Hamiltonian}
\label{AppA}

We analyze in this appendix how to obtain the ladder structure that describes the light-matter interaction int he waveguide. Introducing the creation $a^\dagger_{k}$ and annihilation $a_{k}$ operators for the photons at quasi-momentum $k$ with the CROW, the bath Hamiltonian $H_\text{B}$ can be written as
\begin{equation}
H_\text{B}=\sum_k \omega(k)a^\dagger_{k}a_{k}.
\end{equation}
The waveguide is realized by means of coupled optical resonators separated by a distance 
$h_0$. In this case the band dispersion relation is given by Eq.\ref{dispersion}. 
The coupling between the atomic levels and the electromagnetic field in the CROW is due to a dipole strength $d_{12}$ and characterized by a non-dispersive coupling constant \cite{tudela2014,hung2013}
\bea
g(k)\approx g=g(k_0)= \sqrt{\frac{1}{2\hbar \omega(k_0) \epsilon_0 h_0 M A_m}}\omega_0d_{12},
\eea
where $A_m$ is an effective area that depends on the localization of the mode existing in the other two directions of the waveguide. The resonant wavevector $k_0$ is defined by the relation $k_0=h_0^{-1}\arccos[(\omega_\text{S}-A)/B]$  
\cite{douglas2013,tudela2014,caneva2015,devega2014a}. Hence, the interaction Hamiltonian between the atomic and the photonic degrees of freedom within the rotating wave approximation can be expressed as 
\begin{equation}
H_\text{I}=g\sum_{q=-M/2+1}^{M/2}\sum_{j=1}^{N}\bigg(a_{q} \sigma_j^+ \text{e}^{\text{i}\frac{2\pi q r_j}{h_0 M}}+a^\dagger_{q} 
\sigma_j \text{e}^{-\text{i}\frac{2\pi q r_j}{h_0 M}}\bigg).
\label{hEM3}
 \end{equation}
Finally, we assume that the atoms are located at the position of the optical resonator, i.e.~$r_j= j h_0$ and, therefore, in~\eqref{hEM3} we can use for the photon operators 
the representation in position space
\begin{equation}
a_{j} = \frac{1}{\sqrt{M}} \sum_{q=-M/2+1}^{M/2} a_{q} \text{e}^{\text{i} \frac{2\pi q j}{M}}. 
\label{Fourier}
\end{equation}
Inserting~\eqref{Fourier} in~\eqref{hEM3} and considering the dispersion 
relation~\eqref{dispersion}, we get the Hamiltonian (\ref{DMRG_H}).

\section{Markovian dissipative rates}
\label{AppB}

In order to obtain the Markovian dissipative rates, we consider the long-time limit of~\eqref{rates}. In addition, we take the continuous limit within the momentum sum in the definition of the correlation function~\eqref{Alpha}, such that the Markov dissipative rates can be rewritten as 
		\begin{equation}
			\Gamma_{ij} = \int_{0}^{\infty} \ud t \;\text{e}^{\text{i}  \omega_\text{S} t} \int^{\infty}_{-\infty} \ud k \;\text{e}^{-\text{i}  \omega(k) t} \text{e}^{-\text{i}  k r_{ij}}.
		\end{equation}		 
		Let us now define $\Delta_{k} = \omega(k) - \omega_\text{S}$ and perform the time integral
		\begin{equation}
			\Gamma_{ij} = \lim_{\epsilon \to 0^{+}} \int^{\infty}_{-\infty} \ud k \frac{\text{e}^{-\text{i}  k r_{ij}} }{\epsilon + \text{i}  \Delta_{k}}.
			\label{rate_2}
		\end{equation}		
		For simplicity, we perform the approximation of expanding the cosine around $\pi$ in the dispersion relation of the electromagnetic field
		\begin{equation}
			\omega(k) = A + B\cos(h_0k) \approx A + B \left[ -1 + \frac{h_0^2(k-\pi/h_0)^{2}}{2} \right].
		\end{equation}
Considering in addition the change of variables $k' = k - \pi/h_0$, and defining $a=\frac{2}{Bh_0^2} \left[ \omega_\text{S} - (A-B) + \text{i}  \epsilon \right]$, we can rewrite~\eqref{rate_2} as
		\begin{equation}
			\Gamma_{ij} = \frac{2\text{e}^{-\text{i} \frac{\pi r_{ij}}{h_0}}}{iB}\lim_{\epsilon \to 0^{+}} \int^{\infty}_{-\infty}  \ud k'  \frac{\text{e}^{-\text{i} k' r_{ij}}}{k'^{2} - a}.
		\end{equation}
		We now extend the integral to the complex plane, such that 
\bea
\Gamma_{ij}= \frac{2\text{e}^{-\text{i} \frac{\pi r_{ij}}{h_0}}}{\text{i}  B}\lim_{\epsilon \to 0^{+}} 
\int_{-\infty}^{\infty} \ud z  \frac{\text{e}^{-\text{i}  z r_{ij}}}{(z + \sqrt{a})(z - \sqrt{a})} \nonumber \\
\eea
which gives the approximate result in~\eqref{periodicGamma}.

\section{Time-delayed $\delta$-correlated function}
\label{delta_correlated}

We now show that the time-shifted $\delta$-correlation function given by~\eqref{delta_correlated_correlation} can be obtained by considering the Lorentzian spectrum that is found for light within a cavity \cite{ma2012}. To this aim, we consider a linear dispersion relation for the light $\omega=v_\text{g} |k|$, so that the density of states is $D(\omega)=1/v_\text{g}$. In addition, we now consider in the light-matter interaction Hamiltonian (\ref{hEM3}) the general case in which the coupling strengths are not constant with the momentum, i.e. $g(k)\neq g$. In this situation, the normalized correlation function is no longer written as~\eqref{AlphaC}, but rather as
\bea
\alpha_{nj}(t)=g^{-2}\int_{-\infty}^\infty \ud\omega\;J(\omega)\text{e}^{-\text{i}\omega (t-t^{nj}_\text{d})}, 
\label{limits}
\eea
in terms of the spectral density $J(\omega)=D(\omega)g^2(\omega)$. Also, we have defined an averaged coupling strength $g=\sqrt{\sum_k g_k^2}$, and a delay time $t_\text{d}^{nj}=r_{nj}/v_\text{g}$, such that  $kr_{nj}=\omega t_\text{d}^{nj}$. We now take as in \cite{ma2012} a Lorentzian spectrum centered at $\omega_0$, with height $\beta$ and width $\gamma$
\bea
J(\omega)=\frac{\beta}{\pi}\frac{(\gamma/2)}{(\omega-\omega_0)^2+\big(\frac{\gamma}{2}\big)^2}.
\eea
In this case, we find that $\alpha_{nj}(t)\approx  \beta \text{e}^{-(\gamma+\text{i}\omega_0) (t-t^{nj}_\text{d})}$. Additionally, in the limit of very large Lorentzian width $\gamma\rightarrow\infty$, and considering $\beta=g^2$, we obtain the time-retarded $\delta$-correlated correlation function~\eqref{delta_correlated_correlation}. However, in~\eqref{delta_correlated_correlation} we have not considered the phase with $\omega_0$ since it is not relevant for the discussion.

\section{On the single excitation time evolution and the atom-photon bound states}
\label{AppBound}

When we restrict to the single excitation sector of the many-body Hamiltonian Eq. \eqref{DMRG_H} a number of semi-analytical results concerning the spectrum and the evolution of the system can be obtained.
In this case, for completeness and since it is used in the main text we focus on the case of two 
emitters, for which any wave function can be written as
\begin{align}
|\Psi_t\rangle=(c_1|10\rangle+c_2|01\rangle)\otimes|\textmd{vac}\rangle+\sum_k b_k |001_k\rangle.
\label{2atomstateApp}
\end{align}
By using the fact that the band dispersion relation is symmetric 
in the quasi-momentum one finds that the evolution of $c_\pm=c_1\pm c_2$ are decoupled
and the eigenstates of the systems are given 
by solving the eigenenergy equations
\begin{eqnarray}
 \epsilon^\pm_g-\omega_s=g^2\frac{1\pm \cos(k L)}{ \epsilon^\pm_g-\omega_k+i0^+}.
\label{2atomstateApp}
\end{eqnarray}
In particular in anology with the single atom case discussed in Eq.~\eqref{bounden}
there will be in general two localised atom-photon bound states with energies $\epsilon^\pm_g$ given by the $\pm$ versions of (\ref{2atomstateApp}), respectively.
As discussed in the main text the bound states are very relevant in determining the long time dynamics of the population and entanglement of the emitters.

\section{Analysis of the different scales involved in the atomic evolution}
\label{AppD}

In order to visualize in more detail the different behaviors discussed in Sec. \ref{variation},  
we extract the main time scales of evolution for all the cases analyzed by considering the following general fitting function for the atomic population $P_{\ttot}(t)$:
		\begin{equation}\label{fit}
			F(t) = (I - s_1)\cos(at^{b})\text{e}^{-\lambda_{1} t} + s_2 \left( \text{e}^{-\lambda_{2} t} - 1 \right) + s_1.
		\end{equation}
Here, $I$ is the initial value considered in each case, $P^{ss}_\ttot =F(t\rightarrow \infty) =s_1 - s_2$ represents the steady state, and $a$ and $b$ are fitting parameters that describe the complex oscillatory behavior of the decay. Moreover, the fitting function contains two decaying time scales: a first rate $\lambda_{1}$ that is present along the whole relaxation and can be interpreted as the damping of the oscillations, and a second decaying rate $\lambda_{2}$ that determines the relaxation towards the steady state. Thus, we are now able to characterize the time evolution of the atomic population with three basic parameters: the steady state $P^{ss}_{\ttot}$ and the two decaying rates $\lambda_{1}$ and $\lambda_{2}$.
For more details on the fitting analysis see Appendix \ref{AppC}.

\begin{figure}[!ht]
				\includegraphics[width=1\linewidth]{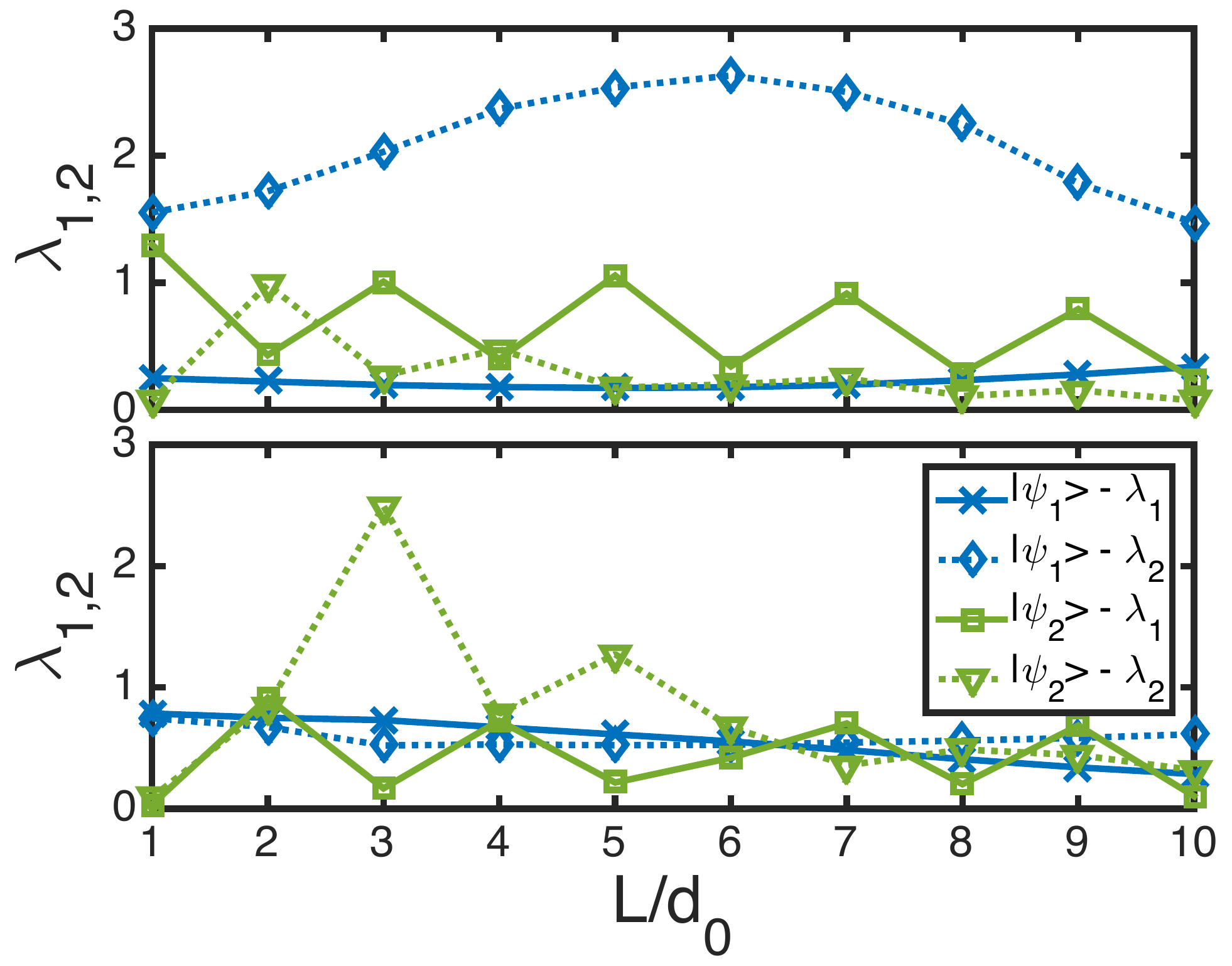}	
\caption{Two panels representing the values of the two decaying scales for only two different conditions $|\psi_1\rangle$ and $|\psi_2\rangle$ with the inter-atomic distance. We have considered $\omega_\text{S}=49$ (top panel) and $\omega_\text{S}=51$ (bottom panel).}
\label{fig:Lambdas}
\end{figure}

\begin{figure}[!ht]
		\includegraphics[width=1\linewidth]{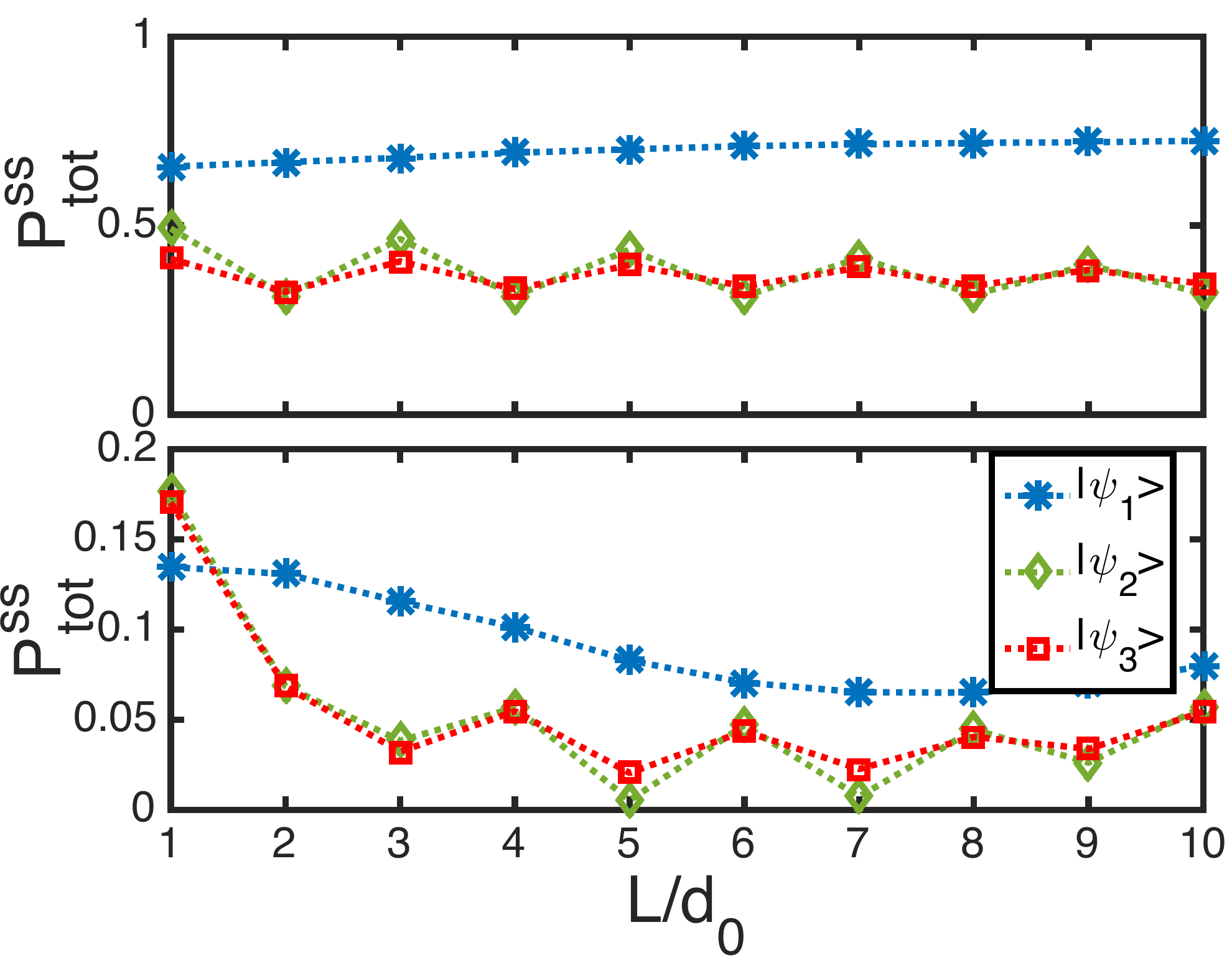}	
\caption{(Color online) Two panels giving the total population of the steady state $P^{SS}_\ttot$ corresponding to $s_1-s_2$ in~\eqref{fit} for the three different initial conditions $|\psi_j\rangle$ for $j=1,2,3$ and corresponding to different inter-atomic distances. We have considered $\omega_\text{S}=49$ (top panel) and $\omega_\text{S}=51$ (bottom panel). }
\label{fig:SteadyStateLambdas}
\end{figure}
\section{Fitting analysis}
\label{AppC}

The accuracy of the fit of~\eqref{fit} has been confirmed when comparing it to other fit functions, as well as by means of a statistical analysis. 
For instance, we considered~\eqref{fit} without the term $ s_2 \left( \text{e}^{\lambda_{2} t} - 1 \right)$ and we analyzed the values of the root-mean-square deviation RMSD, defined as
		\begin{equation}
			\sigma = \sqrt{\frac{\sum_{i} \left( F(x_{i}) - y_{i}\right)^{2}}{N-p}}
		\end{equation}
		where $F$ is the fitting function calculated at the points $x_{i}$, for which the corresponding numerical value is $y_{i}$, $N$ is the number of numerical points, and $p$ is the number of parameters in the fitting function. The value of $\sigma$ for the function without the second exponential decay is always ten times the value of the $\sigma$ for~\eqref{fit}. The latter was of the order of $10^{-3}$, except for some cases in which it reached values of the order $10^{-2}$.

%
Fig.~\ref{fig:Lambdas} represents the value of $\lambda_1$ and $\lambda_2$ for two different initial states $|\psi_1\rangle$ and $|\psi_2\rangle$, and considering the atomic frequencies in the gap and in the band. For the initial state $|\psi_1\rangle$ and $\omega_\text{S}$ within the gap, the most interesting feature is the presence of a large $\lambda_2$, which describes a rapid decay of the system to its steady state value, combined with a much smaller decaying scale $\lambda_1$, which indicates the presence of a very slow damping for the Rabi oscillations around that value. In contrast, both decaying and damping scales are equally dominant (and independent of the relative position of the atoms) when the atomic frequency is within the band. In addition, Fig.~\ref{fig:SteadyStateLambdas} shows that the steady state presents a similar periodic behavior, which is again particularly evident for states containing initial coherences, $|\psi_2\rangle$ and $|\psi_3\rangle$, while for the initial state $|\psi_1\rangle$ is almost negligible.

\section{Accuracy of the Dicke approximation}
\label{Dicke_sec}

In some cases, it is convenient to further simplify the Hamiltonian~\eqref{hEM3} by considering the Dicke approximation such that \cite{dicke1954}
		\bea
		H_\text{I}&=&g\int \ud k \sum_{j=1}^{N}\bigg(a_{k} \sigma_j^+ +a^\dagger_{k} \sigma_j \bigg),
		\label{Dicke}
		\eea
where it is assumed that all the atoms are coupled to the field with the same strength, or in other words that $\text{e}^{-\text{i} k r_j}\approx 1$. In the standard quantum optical case where the atoms are coupled to the radiation field within the vacuum, the Dicke approximation can be safely considered in the limit of $|k_0|L\ll 1$, with $k_0=1/\xi$ the resonant wavevector
of the electromagnetic field and $L$ is the inter-atomic distance. Nevertheless, we have seen that the performance of the weak-coupling approximation is strongly modified by the presence of boundary conditions that give rise to dramatic modifications in the photonic density of states. As a consequence, atoms coupled to such radiation field show very distinct dynamics with respect to the vacuum case. Hence, we may expect that the regime of validity of the Dicke approximation (and therefore the presence of strong collective effects) is altered too. 

\begin{figure}[!ht]
\includegraphics[width=1\linewidth]{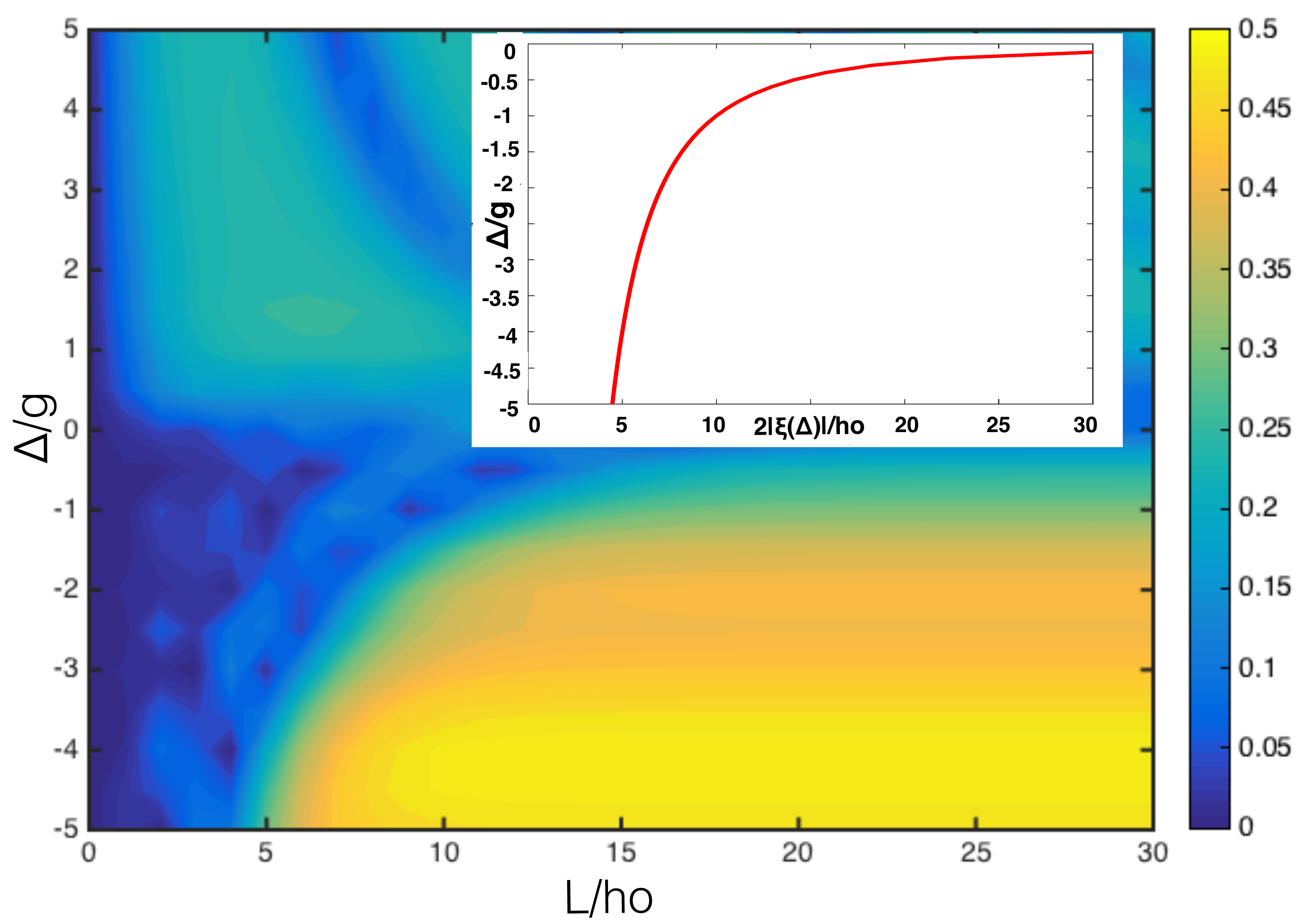}
\caption{(Color online) $E^{\textmd{D}}(L, \omega_\text{S})$ for atom 1. In the contour plot the color scale for the $z$-axis goes from dark blue for zero to bright orange for $0.5$. The $x$-axis holds the values of the distance between the two spins and the $y$-axis the values of $\Delta = \omega_\text{S} - (A-B)$, the difference between the atomic distance and the band-gap edge. The inset represents the line $2|\xi|$ that represents the boundary between distances reached by atom-atom couplings (where collective effects exist) and larger distances where atoms behave as if they were coupled to independent reservoirs.}
\label{fig:dicke}
\end{figure}

\begin{figure}[!ht]
		\includegraphics[width=1\linewidth]{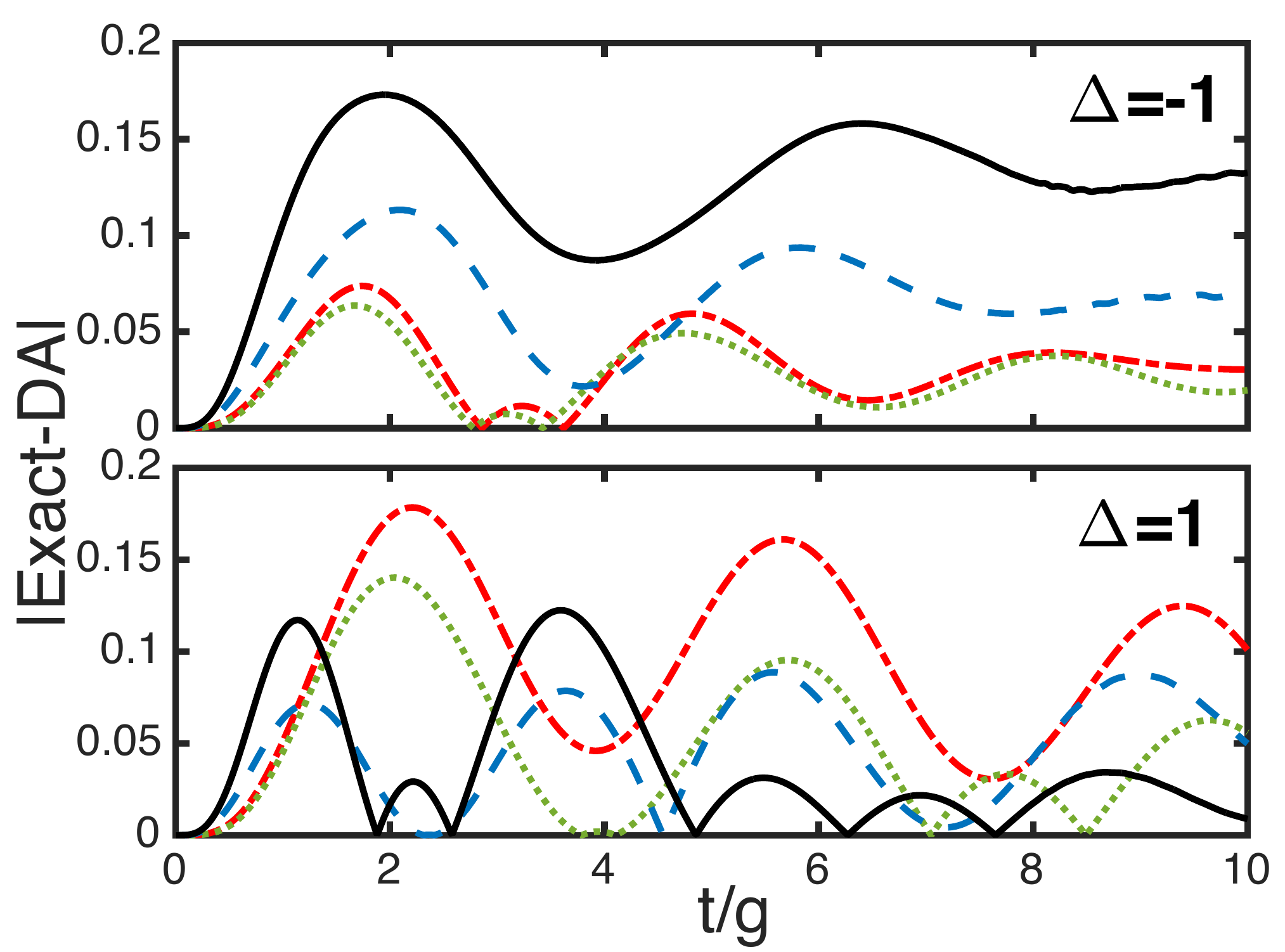}	
\caption{(Color online) Two panels with the difference between the exact population for $L=10$ and the one calculated with the Dicke Hamiltonian (\ref{Dicke}), also following $t$-DMRG. We consider a single atom excited and initial states $|10\rangle$ (solid blue) and $|100\rangle$ (dashed green); and as many excitations as atoms, with initial state $|11\rangle$ (dotted blue), and $|111\rangle$ (dot-dashed green).}
\label{fig:Dicke_2}
\end{figure}
To analyze this, we consider again two atoms with a varying relative distance $L$ and resonant frequencies $\omega_\text{S}$. In addition, we assume as initial condition $|\psi_0\rangle=|10\rangle$. Similarly as in Sec.~\ref{inter-atomic} for the analysis of the accuracy of the ME, we analyze in Fig.~\ref{fig:dicke} the time average of the difference between the result obtained by considering the Dicke approximated Hamiltonian and the original Hamiltonian (\ref{hEM3}), 
\begin{equation}
			E^{\textmd{D}}(L, \omega_\text{S}) = \frac{1}{T} \int_{0}^{T} \ud t \left| P^{\textmd{Dicke}}_{1}(t,L, \omega_\text{S}) - P^{\textmd{Ex}}_{1}(t,L, \omega_\text{S}) \right|
		\end{equation}
where now $P^\text{Dicke}_{1}(t,L, \omega_\text{S})$ and $P^\text{Ex}_{1}(t,L, \omega_\text{S})$ correspond, respectively, to the population of the initially excited atom as computed with the Dicke Hamiltonian~\eqref{Dicke} and the original one given by~\eqref{hEM3}.

It is found that at very short distances $L\lesssim1$ the Dicke approximation is indeed valid for any value of $\omega_\text{S}$. Besides this case the Dicke approximation is also accurate in two situations: in the band-gap edge and deep in the band for certain periodic values of the atomic separation. In contrast, the Dicke approximation performs very poorly for atomic frequencies within the gap. 

These results can be qualitatively explained in the Markov limit by considering the behavior of the dissipative rates~\eqref{periodicGamma}. Indeed, in the band-gap edge the length of the atom-atom interactions $\xi$ is infinite, which means that all atoms are equally connected with each other through the field, as is very well-described by the Hamiltonian~\eqref{Dicke}. Also, for frequencies within the band the value of the atom-atom rates is periodic with a long period $\lambda_L=d_0\pi\sqrt{B/2\Delta}$. Hence, the validity of the Dicke approximation also displays such periodicity, particularly in frequency regions well inside the band where the Markovian-approximated rates~\eqref{periodicGamma} describe accurately the dynamics. Finally, within the gap the Dicke approximation works at distances that are smaller than the minimum distance at which adjacent atoms are connected, i.e.~$L<2|\xi|$ (see this line in the inset of Fig.~\ref{fig:Dicke_2}). Such a line corresponds perfectly with the separation between the blue region where the Dicke approximation works and the yellow region where it is not a good approximation because it overestimates the presence of collective effects.

To analyze the influence of the presence of more than one excitation in the system, we display in Fig.~\ref{fig:Dicke_2} the evolution of a quantity similar to~\eqref{errorMEE}, 
\bea
E(t) = \frac{1}{t} \int_{0}^{t} \ud s \left| P^{\textmd{Dicke}}_\ttot(s) - P^{\textmd{total}}_{\ttot}(s) \right|.
\label{timeevolerror_2}
\eea
where now $P^{\textmd{Dicke}}_\ttot=N^{-1}_E\sum_{j=1}^NP^{\textmd{Dicke}}_{1}$ computed with~\eqref{Dicke}, and $P^{\textmd{total}}_\ttot$ is the same quantity computed with the non-approximated Hamiltonian~\eqref{hEM3}, with both cases calculated in $t$-DMRG. Here, the total population is divided by the total number of excitations in the problem, $N_E$, and we consider $N_E=1, 2, 3$. 
As can be seen, the error between the prediction from the Dicke approximation and the exact Hamiltonian is only sensitive to the number of existing excitations when the atomic frequencies are in the gap. Hence, the inability of the Dicke approximation to describe correctly the atom-photon bound state is even more dramatic when more excitations are present.

\section{Exact Diagonalization and DMRG}
\label{ed_dmrg_sec}

The Exact Diagonalization is the most straightforward numerical method, as it consists in evolving the initial state of the system by applying the unitary time evolution operator $U = e^{-\imath H t}$, where $H$ is the full Hamiltonian of the system. 
Since the dimension of the Hilbert space grows with the number of excitations allowed in the system and with the number of oscillators considered to represent the electromagnetic field, this method has several numerical limitations and it is not suitable to simulate very complex configurations. 
In detail, the dimension of the Hilbert space is given by 
\begin{equation}
	N = dim(\mathcal{H}) = \sum_{i=0}^{N_{a}} \sum_{j=i}^{N_{e}} \dbinom{N_{a}}{i} \dbinom{N_{o} + j - i - 1}{j - i}
\end{equation}
with $N_{a}$ number of spins, $N_{o}$ number of oscillators (bosons) and $N_{e}$ maximum number of excitations in the system. In our case, we use the ED method for a single excitation, by projecting the full Hamiltonian $H$ in the basis ${\mathcal B}=\{|\textmd{vac}\rangle,|1_j\rangle\}$, where now $|\textmd{vac}\rangle$ represents no excitations neither in the atoms nor in the bath, and $|1_j\rangle$ represents a single excitation at the site $j$ representing either atomic or bath observables. Projected in this basis, the size of $H$ grows only linearly with the number of atoms and environment oscillators, which leads to a short computational time even when considering several hundreds oscillators in the EM field. However, the number of field oscillators that can be numerically implemented when considering two excitations decreased dramatically, compromising compliance with the requirement that the bath should be sufficiently large so as to produce full atomic relaxation without being affected by finite size effects. 

With such a limitation either an approximated method such as the weak coupling ME or more performant numerical methods like $t$-DMRG are needed to solve the dynamics of the open system with more than one excitation. To implement $t$-DMRG we used the toolkit by Ian McCulloch \cite{IanToolkit} that is based on representing the full system wave vector as a matrix product state of the form 
\bea
|\Psi\rangle&=&\sum_{i_1,\cdots,i_N} A[1]^{i_1}A[2]^{i_2}\cdots A[N-1]^{i_{N-1}}A[N]^{i_{N}}\cr
&\times&|i_1,i_2,\cdots,i_N\rangle.
\eea
Here $|i_j\rangle$ represents the local basis of the particle site $i$, which can either be an atom or a harmonic oscillator, and $A[k]^i$ represent a set of matrices for each particle site with a dimension (the bond dimension) that is larger the more entanglement needs to be encoded. 

The ladder structure (\ref{DMRG_H}) in which we have mapped the system Hamiltonian contains only local couplings, which ensures that at low energies the entanglement will grow polynomially and not exponentially with the number of degrees of freedom, and therefore that the required bond dimension is small \cite{eisert2008}. However, because of the presence of the spins in between the bosons, our Hamiltonian produces interactions that are not only between nearest-neighbor sites, but also between next-nearer neighbors. For this reason, the time evolution can not be performed with the standard Trotter-Suzuki algorithm, but rather with a more sophisticated Krylov method. This is based on using a Lanczos technique to dynamically generate a basis of the most relevant states explored during the evolution \cite{LanczosTime}. This evolution method was also checked with time evolving block decimation \cite{Vidal2003,Verstraete2004,Zwolak2004}. The convergence tests performed have lead us to set a truncation error bound per time step of $10^{-5}$, a timestep of $\Delta t = 0.01$ and a maximum number of Krylov vectors per step of $15$.

\bibliography{Bibtexelesdrop2}

\end{document}